\documentclass[
  reprint,
  amsmath,
  amssymb,
  aps,
  prapplied,
  floatfix,
  showkeys,
]{revtex4-2}

\usepackage{array}[=2016-10-06]

\usepackage{graphicx} %
\usepackage[caption=false]{subfig}
\usepackage[export]{adjustbox}

\usepackage{dcolumn}
\usepackage{tabularx}

\usepackage{bm}

\usepackage{mathtools} %

\usepackage{physics2}  %
\usephysicsmodule{ab, braket}

\usepackage{derivative}

\usepackage{siunitx}

\usepackage{xcolor}  %
\usepackage{mdframed}

\usepackage[section]{placeins}

\usepackage{hyperref}  %
\newtheorem{theorem}{Theorem}

\DeclareMathOperator{\Tr}{Tr}
\DeclareMathOperator{\vecop}{vec}

\NewDocumentCommand{\proj}{m}{\ketbra{#1}{#1}}

\DeclarePairedDelimiterX{\flsket}[1]{\lvert}{\rangle\!\rangle}{#1}

\begin{document}

\title{The Norton Theorem for Quantum Circuits and Systems}

\author{Anthony J. Cressman}
\email{anthony.j.cressman.gr@dartmouth.edu}
\affiliation{
  Department of Physics \\
  Dartmouth College \\
  Hanover, NH, USA
}

\author{Rahul Sarpeshkar}
\email{rahul.sarpeshkar@dartmouth.edu}
\affiliation{
  Departments of Engineering, Physics, Microbiology \& Immunology, and Molecular \& Systems Biology \\
  Dartmouth College \\
  Hanover, NH, USA
}

\date{\today}

\begin{abstract}
  Previous work has shown that quantum \(N \times N\) Hamiltonians with complex variables can be exactly mapped to \(2N \times 2N\) classical analog circuit equivalents with real and imaginary variables.
  Therefore, complex quantum circuits and computations, including NMR, Josephson junctions, the quantum Fourier transform, and density matrix evolution, can be implemented exactly on analog VLSI chips and circuits, including loss.

  It is well known that classical analog circuits and systems benefit from powerful Th\'evenin and Norton theorems that rigorously and exactly reduce the rest of a network to equivalent sources and impedances driving a selected subsystem.
  Here, we construct single-port and multiport Norton equivalent circuits that reproduce the dynamics of selected coordinates of a quantum system.
  We partition the Schr\"odinger equation into retained and eliminated sectors and apply Gaussian elimination to yield an exact reduced equation wherein the eliminated sector contributes a self-energy term \(\Sigma(s)\), which is realized as a Norton admittance, and an effective source term \(F(s)\), realized as a Norton current source.
  In multiport settings, the reduction is the Schur complement of the eliminated block.
  The same construction applies to composite system-environment problems once the product basis is partitioned into retained and eliminated configuration subspaces; for finite closed systems, the reduction is exact.
  The framework works for closed system state vector dynamics, density matrix dynamics, and open system evolution governed by the Lindblad equation, with a time domain extension for time-dependent Hamiltonians.

  We demonstrate the framework with Grover search, where the Norton reduction yields a single bright pole in the ideal case and diagonal disorder redistributes spectral weight into dark poles and degrades performance.
  The corresponding circuit implementations reproduce the exact reduced dynamics, and a renormalized bright pole model remains accurate through a weak-to-moderate disorder regime before the dark sector significantly affects the response.
  Therefore, the quantum Norton theorem can help design and analyze quantum circuits and computers as powerful extensions of analog circuits and computers, providing a new theoretical, conceptual, and experimental circuit tool.
\end{abstract}


\maketitle

\section{\label{sec:intro} Introduction}

\begin{figure*}
  \centering
  \subfloat[]{
    \includegraphics[height=0.1\textheight]{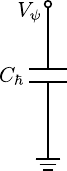}
    \label{fig:circuit_symbols:complex_planck_capacitor}
  }\hfill
  \subfloat[]{
    \includegraphics[height=0.1\textheight]{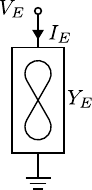}
    \label{fig:circuit_symbols:complex_qu_admittance}
  }\hfill
  \subfloat[]{
    \includegraphics[height=0.1\textheight]{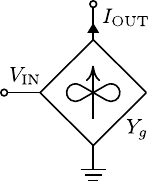}
    \label{fig:circuit_symbols:complex_qu_transadmittance}
  }\hfill
  \subfloat[]{
    \includegraphics[height=0.1\textheight]{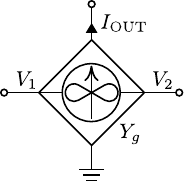}
    \label{fig:circuit_symbols:complex_qu_transadmittance_mixer}
  }

  \caption{
    Primitive circuit elements used throughout the paper to represent quantum circuit implementations.
    For compactness, all symbols are drawn in complex form, with each drawn ``complex'' wire representing two underlying real signal channels, one for the real part and one for the imaginary part of the signal~\cite{sarpeshkarEmulationQuantumQuantuminspired2019,sarpeshkarEmulationQuantumQuantuminspired2019a,sarpeshkarEmulationQuantumQuantuminspired2020,sarpeshkarQuantumCochleaEfficient2019,cressmanSarpeshkarFormulationEmulationQuantuminspired2022}.
    (a) A Planck capacitor, whose voltage represents a quantum state variable and whose capacitance sets the \(\hbar\) scale of the dynamical equation.
    (b) A quantum admittance, implementing a diagonal Hamiltonian term.
    (c) A quantum transadmittance, implementing a constant off-diagonal coupling.
    (d) A quantum transadmittance mixer, implementing time-dependent or controlled couplings.
  }
  \label{fig:circuit_symbols}
\end{figure*}

\begin{figure*}
  \centering
  \includegraphics[max width=\textwidth]{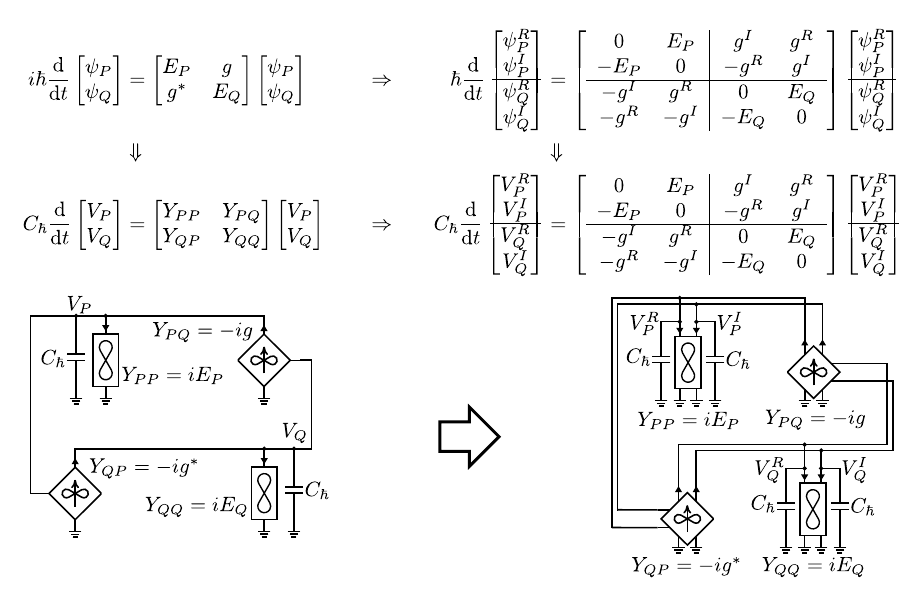}
  \caption{
    Exact equivalence between a two-state Schr\"odinger equation and its circuit representations.
    \emph{Top:} for a two-state system, the complex Schr\"odinger equation is separated into real and imaginary parts, illustrating the general conversion of an \(N\times N\) complex Hamiltonian system into an equivalent \(2N\times2N\) real dynamical system.
    \emph{Middle:} the same dynamics are written first as a compact complex circuit equation and then as a fully real circuit equation by resolving real and imaginary components.
    \emph{Bottom left:} compact ``complex'' circuit representation, with each wire representing two channels for the real and imaginary components.
    \emph{Bottom right:} fully realized ``real'' circuit representation, with both the real and imaginary signals shown.
    In the compact form, the Hamiltonian \(H\) is encoded as a complex admittance matrix \(Y\), with \(Y=-iH\) under the chosen convention.
    In the expanded form, the same dynamics appear as real couplings between the real and imaginary components, realized with Planck capacitors and real-valued voltage-controlled current sources.
  }
  \label{fig:tls_mapping}
\end{figure*}

\begin{figure}
  \centering
  \includegraphics[max width=\columnwidth]{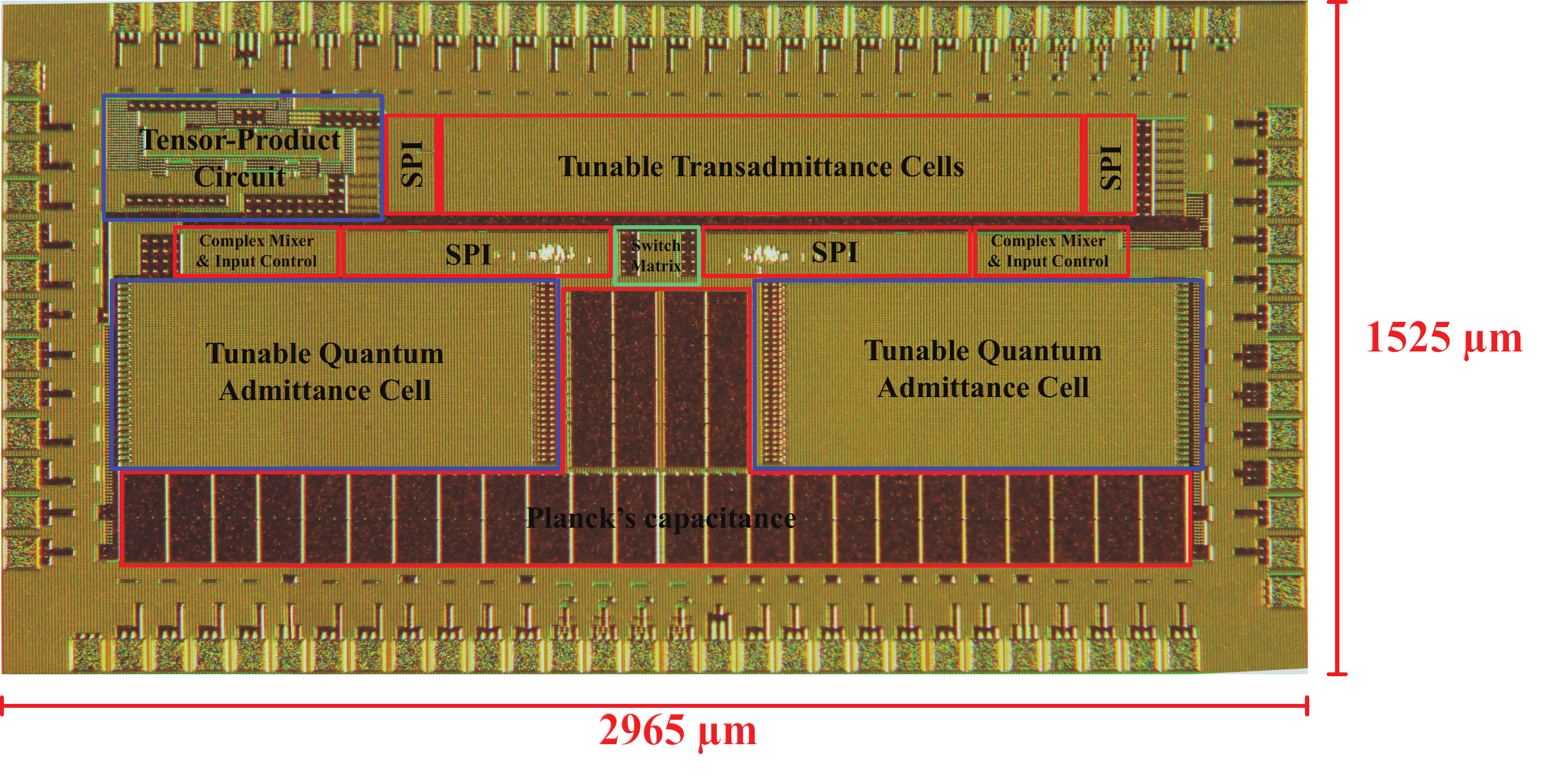}
  \medskip
  \includegraphics[max width=\columnwidth]{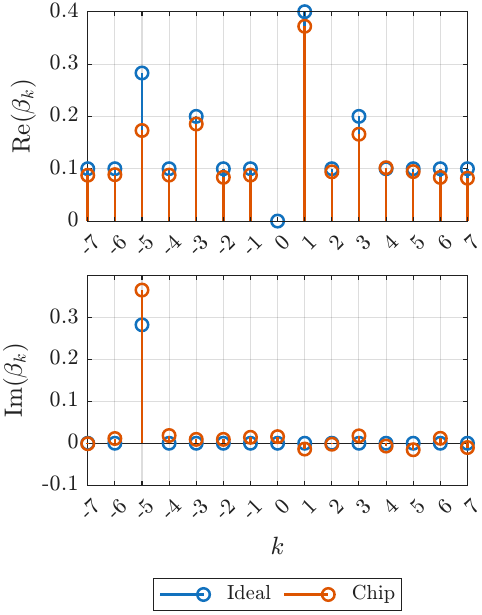}

  \caption{
    Prior experimental silicon realization of the quantum-inspired circuit framework of Ref.~\cite{cressmanSarpeshkarFormulationEmulationQuantuminspired2022}.
    \emph{Top:} fabricated \(\qty{0.18}{\micro\meter}\) analog integrated circuit implementing the core programmable circuit primitives.
    \emph{Bottom:} representative quantum Fourier transform (QFT) readout from the same hardware platform, comparing ideal coefficients with measured chip outputs for \(\mathrm{Re}(\beta_k)\) and \(\mathrm{Im}(\beta_k)\).
  }
  \label{fig:chip_results}
\end{figure}

\begin{figure*}
  \centering
  \includegraphics[max width=\textwidth]{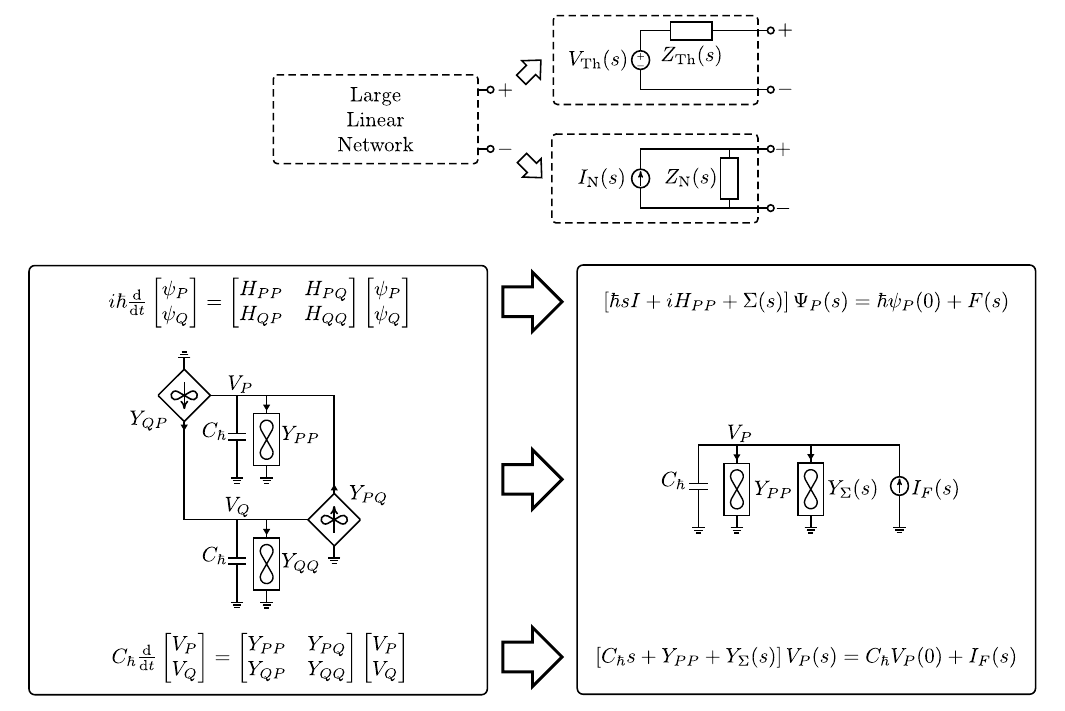}
  \caption{
    \emph{Top:} \textbf{Classical reduction of a linear network to equivalent single-port descriptions in the Laplace domain.}
    \emph{Left:} a linear network viewed from the terminal pair \(+\) and \(-\).
    \emph{Upper right:} the Th\'evenin equivalent, represented by an effective voltage source \(V_{\mathrm{Th}}(s)\) in series with an impedance \(Z_{\mathrm{Th}}(s)\).
    \emph{Lower right:} the Norton equivalent, represented by an effective current source \(I_{\mathrm{N}}(s)\) in parallel with an equivalent impedance \(Z_{\mathrm{N}}(s)\), or equivalently an admittance \(Y_{\mathrm{N}}(s)=1/Z_{\mathrm{N}}(s)\).
    The dependence on \(s\) indicates that these equivalent sources and impedances are generally frequency-dependent.
    For a linear single-port network, \(Z_{\mathrm{N}}(s)=Z_{\mathrm{Th}}(s)\), so the original network, its Th\'evenin equivalent, and its Norton equivalent have identical terminal behavior~\cite{siebertCircuitsSignalsSystems1986}.
    \emph{Bottom:}
    \textbf{Reduction of a coupled quantum system to a Norton-equivalent representation.}
    \emph{Left:}
    A partitioned quantum system with retained (\(P\)) and eliminated (\(Q\)) coordinates is described by coupled Schr\"odinger equations and an equivalent circuit with self-admittances and transadmittances.
    \emph{Right:}
    Eliminating the \(Q\) subsystem via a Schur complement yields an exact reduced description involving only the retained coordinates.
    The influence of the eliminated subsystem appears as a frequency-dependent self-energy \(\Sigma(s)\) and an effective driving term \(F(s)\).
    In the circuit representation, this corresponds to a compact Norton equivalent consisting of an effective admittance \(Y_\Sigma(s)\) in parallel with a current source \(I_F(s)\).
  }
  \label{fig:overview}
\end{figure*}

Classical physical systems, and especially classical analog circuits, have been used to reproduce certain aspects of quantum system dynamics and quantum information processing.
Coupled classical oscillators can recreate the resonant transfer underlying quantum search~\cite{groverSenguptaClassicalAnalogQuantum2002}, and recent work on classical oscillator networks has developed quantum-like product states~\cite{scholesAmatiQuantumlikeProductStates2025} and gate operations~\cite{amatiScholesQuantumInformationQuantumlike2025} with applications to simulating spin system dynamics~\cite{sahaScholesQuantumDynamicsSpin2026}.
Classical analog circuits and signal processing approaches have been used for simulating quantum states as well as quantum operators on those states~\cite{kishQuantumComputingAnalog2003,lacourOttSignalbasedClassicalEmulation2015}.
Other classical circuit architectures have been used to reproduce quantum walks~\cite{ezawaElectriccircuitSimulationSchrodinger2019} and features of topological quantum computing~\cite{zouZhangExperimentalSimulationTopological2023,chenZhangEngineeringTopologicalStates2025}.

Recently, classical analog circuits have been used to model Schr\"odinger dynamics, Josephson junctions, the quantum Fourier transform, and the dynamics of density matrix states~\cite{sarpeshkarEmulationQuantumQuantuminspired2019,sarpeshkarEmulationQuantumQuantuminspired2019a,sarpeshkarEmulationQuantumQuantuminspired2020,sarpeshkarQuantumCochleaEfficient2019,cressmanSarpeshkarFormulationEmulationQuantuminspired2022,cressmanSarpeshkarEmulationDensityMatrix2025}.
The framework used here maps an \(N \times N\) complex Hamiltonian system exactly to a \(2N \times 2N\) paired-real analog circuit system.
In this mapping, each probability amplitude is represented by a pair of voltages (for the real and imaginary parts) on Planck capacitors, and Hamiltonian terms are implemented by the primitive circuit elements shown in Fig.~\ref{fig:circuit_symbols}: quantum admittances, quantum transadmittances, and quantum transadmittance mixers.
Figure~\ref{fig:tls_mapping} shows how these primitives assemble into the equivalent circuit descriptions for a two-state Hamiltonian.
These primitive elements have been experimentally realized in silicon and used to emulate nuclear magnetic resonance (NMR), Josephson junction, and quantum Fourier transform dynamics (Fig.~\ref{fig:chip_results})~\cite{cressmanSarpeshkarFormulationEmulationQuantuminspired2022}.
The framework has also been extended for simulating density matrix dynamics, providing a path for simulating mixed states, noisy and dissipative states, and general open quantum systems~\cite{cressmanSarpeshkarEmulationDensityMatrix2025}.

It is frequently the case that only a few degrees of freedom of a full quantum state are desired to be simulated (e.g., separating the dynamics between a system and an environment)~\cite{breuerPetruccioneTheoryOpenQuantum2010,manzanoShortIntroductionLindblad2020,chinPlenioExactMappingSystemreservoir2010}.
Similarly, it is often desirable to replace large, complex analog circuit networks with simpler equivalents, without changing the behavior at a chosen set of terminals.
The mathematical frameworks for these types of model reduction are well established in both quantum mechanics and electrical engineering.
In Hilbert space, projection methods such as Feshbach projection and L\"owdin partitioning obtain effective Hamiltonians for selected subspaces, and Green's function methods describe the influence of eliminated degrees of freedom through self-energies~\cite{feshbachUnifiedTheoryNuclear1958,feshbachUnifiedTheoryNuclear1962,lowdinStudiesPerturbationTheory1962,economouGreensFunctionsQuantum2006,dattaQuantumTransportAtom2005,dattaElectronicTransportMesoscopic2009}.
In Liouville space, the Nakajima-Zwanzig projection formalism~\cite{nakajimaQuantumTheoryTransport1958,zwanzigEnsembleMethodTheory1960} describes reduced density matrix dynamics through memory kernels~\cite{smirneVacchiniNakajimazwanzigTimeconvolutionlessMaster2010} and, in general, initial condition source terms.
In electrical engineering, Kron reduction eliminates internal network nodes to preserve the behavior of a chosen set of terminals~\cite{kronTensorAnalysisNetworks1939,dorflerBulloKronReductionGraphs2013}, while Th\'evenin and Norton equivalents replace networks with equivalent impedances and sources~\cite{siebertCircuitsSignalsSystems1986}.
Many of these reduction methods can be expressed through Gaussian elimination and Schur complements~\cite{golubVanLoanMatrixComputations2013}.

In this paper we formulate a quantum Norton theorem for the modeling of quantum systems with classical analog circuits.
In this construction, a quantum state is partitioned into coordinates to be retained \(P\) and coordinates to be eliminated \(Q\).
For a time-independent Hamiltonian \(H\), the \(Q\) sector is algebraically eliminated in Laplace space with Gaussian elimination or the Schur complement, leaving an exact equation for \(P\) alone.
The new reduced equation includes a frequency-dependent self-energy \(\Sigma(s)\)  and source \(F(s)\), which we realize as a Norton equivalent admittance \(Y_\Sigma(s)\) and independent current source \(I_F(s)\) connected in parallel.
The poles and residues of the admittance determine the auxiliary circuit modes and their couplings, and the current source preserves the effects of the initial conditions of the eliminated sector on the retained state dynamics.
An overview of the mapping and its relation to the classical Norton and Th\'evenin equivalent circuits is presented in Fig.~\ref{fig:overview}.

There are five benefits to this approach:
(1) In applications that require both classical and quantum operations, e.g., in chemistry, we present a rigorous unifying circuits framework that is capable of handling both simultaneously, in widely available circuit design software, including examples of noise, loss, and approximation;
(2) The previously demonstrated VLSI chip, which implements the QFT and tensor-product classically is an example of applied hardware to which this framework can apply.
Details of the power, energy, and part scaling of digital, analog, collective analog and quantum systems are discussed in Appendix~\ref{sec:app:scaling}.
(3) In practical quantum circuit design, as in classical circuit design, a principled way to accumulate all noise sources from various component noisy devices into one effective loss or into one effective noise source is highly useful; the Lindblad extensions of this framework show how this is possible;
(4) Algorithms like Grover's algorithm effectively create `non-marked loser items' that all accumulate into one effective loser strength that attenuates over a period, while the marked `winner' accumulates in strength in this period.
Classical analog winner-take-all circuits~\cite{hahnloserSeungDigitalSelectionAnalogue2000} and the circuit realizations of Grover dynamics presented here demonstrate this behavior as well, motivating the choice of Grover dynamics as an example.
Identical scaling has previously been demonstrated for classical coupled-oscillator and quantum search dynamics~\cite{groverSenguptaClassicalAnalogQuantum2002}.
(5) Probability directly translates to Planck capacitor energies and Planck capacitor power flows directly correspond to energy flows; thus, we can use such mappings to design and analyze quantum systems and even measure these power flows in actual working hardware.
Indeed, the room temperature Josephson junction emulation reported in Ref.~\cite{cressmanSarpeshkarFormulationEmulationQuantuminspired2022} exploited this technique.

In the Norton representation, circuit savings likewise depend on exploitable structure or approximations, as the dynamics of the eliminated sector must still be modeled with auxiliary modes.
However, one can characterize the pole structure and coupling strengths of the eliminated sector to make approximations which reduce the number of required circuit components.
In addition to practical considerations, there are also a number of conceptual advantages to this reduction.
For example, the dynamics of the continuous-time Grover Hamiltonian can be understood as the interaction between a retained/marked sector and a single bright mode representing the eliminated/unmarked sector.
In these examples, the marked state is known at the time of construction, and the purpose is to demonstrate the reduction of its dynamics rather than to find an unknown marked state.
Breaking the symmetry of the ideal Grover Hamiltonian with disorder activates additional modes that appear in the Norton reduction, revealing how changes in the symmetry of the eliminated sector lead to degraded search performance.

The structure of the paper is as follows.
We first review classical Norton/Th\'evenin reduction and the analog circuit representation of quantum dynamics used throughout (Sec.~\ref{sec:circuit_review}).
We develop the single-port quantum Norton theorem (Sec.~\ref{sec:qnt}) and use it to show how the dynamics of the marked state in the typical Grover search Hamiltonian can be recreated with a much simpler reduced circuit (Sec.~\ref{sec:grover}).
The single-port quantum Norton theorem is then generalized to the multiport case (Sec.~\ref{sec:general_qnt}), and the pole/residue structure of the eliminated sector (and approximation thereof) is analyzed through a series of finite and continuum-like examples (Sec.~\ref{sec:pole_structure}).
We then return to the ideal continuous-time Grover search problem and show how the performance of a non-ideal, disordered Grover Hamiltonian can be attributed to changes in the pole structure from a single bright branch to a dominant bright branch with dark sector corrections (Sec.~\ref{sec:disordered_grover}).
Finally, we extend this framework to composite systems, Liouville space dynamics, and dissipative effective Hamiltonians (Sec.~\ref{sec:conditioning}).
The circuit demonstrations are simulated in LTspice, a freely available circuit simulator.
\section{\label{sec:circuit_review} Classical Port Reduction and Quantum Circuit Representation}

\begin{table*}[ht]
  \setlength{\extrarowheight}{4pt}
  \begin{tabularx}{\textwidth}{|X|X|}
    \hline
    \centering\textbf{Quantum mechanics}
     & \centering\textbf{Classical analog circuit emulation} \tabularnewline
    \hline
    Complex probability amplitude
     & Real and imaginary voltage pairs stored on paired Planck capacitors                         \\
    \hline
    Global Planck constant \(\hbar\)
     & A global Planck capacitance \(C_\hbar\) that can be chosen for the electronic chip          \\
    \hline
    Physical time scale set by the system
     & Emulation can be slowed down or sped up by choosing the Planck capacitance                  \\
    \hline
    Thermal excitation depends on quantum energy scales relative to \(k_B T\). Cooling may be needed
     & Capacitor signal energies can be chosen well above \(k_B T\) for room-temperature emulation \\
    \hline
    Energy-based Hamiltonian interactions
     & Current-based implementation of Hamiltonian interactions                                    \\
    \hline
    Probability conservation
     & Conservation of total Planck capacitor energy                                               \\
    \hline
    Energy conservation for time-independent \(H\)
     & Conservation of the corresponding mapped power-flow quantity                                \\
    \hline
  \end{tabularx}
  \caption{Correspondence between quantum mechanics and classical analog circuit emulation~\cite{cressmanSarpeshkarFormulationEmulationQuantuminspired2022}.}
  \label{tab:correspondence}
\end{table*}

\subsection{\label{sec:circuit_review:classical} Classical Norton and Th\'evenin Equivalents}

Large circuit networks are cumbersome to analyze and costly to implement directly when only the behavior at a chosen terminal pair matters.
Classical circuit theory resolves this by replacing the internal network with an equivalent reduced single-port description that preserves the same external behavior.
For a single port, this description takes either Th\'evenin or Norton form~\cite{siebertCircuitsSignalsSystems1986}.
As shown in Fig.~\ref{fig:overview}, a large, complicated circuit network is represented in the Laplace domain by either an effective voltage source in series with an impedance or an effective current source in parallel with an equivalent impedance.
The Th\'evenin and Norton equivalent circuits are exactly equivalent to each other and produce the same terminal voltage-current relation for any external connection at the retained port.
In either case, the equivalent circuit reduction is obtained by algebraically eliminating the internal network variables from the system of circuit node equations in the Laplace domain to produce an equivalent single-port description at the retained terminals.

\subsection{\label{sec:circuit_review:quantum} Quantum Dynamics as Analog Circuits}

Previous work established an exact mapping from the Schr\"odinger equation to an analog circuit network built from a small set of primitive elements~\cite{sarpeshkarEmulationQuantumQuantuminspired2019, sarpeshkarEmulationQuantumQuantuminspired2019a, sarpeshkarEmulationQuantumQuantuminspired2020, sarpeshkarQuantumCochleaEfficient2019,cressmanSarpeshkarFormulationEmulationQuantuminspired2022}.
In this representation, complex quantum amplitudes are encoded by two real voltage channels (one for the real part and one for the imaginary part), and couplings between them are implemented as controlled current flows.
Overall, \(2N = 2^{n+1}\) real-valued signals are required to model general state vector dynamics, though symmetry can often be exploited to reduce this requirement~\cite{cressmanSarpeshkarFormulationEmulationQuantuminspired2022,cressmanSarpeshkarEmulationDensityMatrix2025}.
Fuller treatments of the circuit components necessary for this discussion are given in Refs.~\cite{sarpeshkarEmulationQuantumQuantuminspired2019, sarpeshkarEmulationQuantumQuantuminspired2019a, sarpeshkarEmulationQuantumQuantuminspired2020, sarpeshkarQuantumCochleaEfficient2019, cressmanSarpeshkarFormulationEmulationQuantuminspired2022, cressmanSarpeshkarEmulationDensityMatrix2025}.

Separating the state and Hamiltonian into real and imaginary parts,
\begin{equation}
  \ket{\psi} = \ket{\psi^R} + i\ket{\psi^I},
  \qquad
  H = H^R + iH^I,
\end{equation}
the Schr\"odinger equation
\begin{equation}
  i\hbar \odv{}{t}\ket{\psi} = H\ket{\psi}
\end{equation}
takes the form
\begin{equation}
  \hbar \odv{}{t}
  \begin{bmatrix}
    \psi^R \\
    \psi^I
  \end{bmatrix}
  =
  \begin{bmatrix}
    H^I  & H^R \\
    -H^R & H^I
  \end{bmatrix}
  \begin{bmatrix}
    \psi^R \\
    \psi^I
  \end{bmatrix},
\end{equation}
where we now identify \(\ket{\psi^R}\) and \(\ket{\psi^I}\) with real coordinate vectors \(\psi^R\) and \(\psi^I\), and \(H^R\) and \(H^I\) are real matrices.
Identifying \(V^R \equiv \psi^R\) and \(V^I \equiv \psi^I\), and choosing the capacitance scale \(C_\hbar\) to represent the factor of \(\hbar\), gives the circuit equation
\begin{equation}
  C_\hbar \odv{}{t}
  \begin{bmatrix}
    V^R \\
    V^I
  \end{bmatrix}
  = -\begin{bmatrix}
    Y^R & -Y^I \\
    Y^I & Y^R
  \end{bmatrix}
  \begin{bmatrix}
    V^R \\
    V^I
  \end{bmatrix},
\end{equation}
Here \(Y=Y^R+iY^I\) is the effective complex admittance matrix, so \(C_\hbar\dot V+YV=0\) and \(Y\leftrightarrow iH\) under the chosen circuit mapping.
Each quantum amplitude is represented by a pair of capacitor voltages, with the capacitance scale \(C_\hbar\) corresponding to \(\hbar\), which also determines the speed of the simulated dynamics.
We refer to these as Planck capacitors [Fig.~\ref{fig:circuit_symbols}(a)].
The entries of the Hamiltonian then appear as effective admittances and transadmittances in the circuit network.
Diagonal terms map to local quantum admittances [Fig.~\ref{fig:circuit_symbols}(b)], while off-diagonal terms map to quantum transadmittances [Fig.~\ref{fig:circuit_symbols}(c)].
The quantum transadmittance mixer [Fig.~\ref{fig:circuit_symbols}(d)] implements time-dependent or controlled couplings.

These parts implement arbitrary finite-dimensional quantum dynamics.
For example, consider a two-level system with Hamiltonian
\begin{equation}
  H_{\mathrm{TLS}} =
  \begin{bmatrix}
    E_P & g   \\
    g^* & E_Q
  \end{bmatrix}.
\end{equation}
In the circuit representation, the diagonal terms \(E_P\) and \(E_Q\) appear as local quantum admittances at the two nodes, while the off-diagonal coupling \(g\) appears as a quantum transadmittance coupling between them.
For each state, the real and imaginary components are represented by voltages on separate Planck capacitors.
Figure~\ref{fig:tls_mapping} shows the full circuit obtained from this mapping for a two-level system.
The circuit construction reduces the quantum reduction problem to the classical task of network reduction.
The basic physical correspondences are summarized in Table~\ref{tab:correspondence}.

For Hermitian \(H\), if \(H \ket{E_k} = E_k \ket{E_k}\), then
\begin{equation}
  \begin{aligned}
    \begin{bmatrix}
      H^I  & H^R \\
      -H^R & H^I
    \end{bmatrix}
    \begin{bmatrix}
      \ket{E_k} \\
      -i\ket{E_k}
    \end{bmatrix}
     & =
    -iE_k
    \begin{bmatrix}
      \ket{E_k} \\
      -i\ket{E_k}
    \end{bmatrix},
    \\
    \begin{bmatrix}
      H^I  & H^R \\
      -H^R & H^I
    \end{bmatrix}
    \begin{bmatrix}
      \ket{E_k}^{*} \\
      i\ket{E_k}^{*}
    \end{bmatrix}
     & =
    iE_k
    \begin{bmatrix}
      \ket{E_k}^{*} \\
      i\ket{E_k}^{*}
    \end{bmatrix}.
  \end{aligned}
\end{equation}
Thus, each quantum energy eigenvalue \(E_k\) maps to the conjugate eigenvalue pair \(-iE_k\) and \(+iE_k\), corresponding to dynamical rates \(-iE_k/\hbar\) and \(+iE_k/\hbar\).
The corresponding conjugate eigenvectors combine to produce real voltage trajectories in the circuit implementation.
\section{\label{sec:qnt} The Quantum Norton Theorem}

A large quantum system where the dynamics of only a few coordinates is desired may be similarly reduced as in the large circuit problem.
For a time-independent Hamiltonian, the Schr\"odinger equation is
\begin{equation}
  i\hbar \odv{}{t}\ket{\psi(t)} = H \ket{\psi(t)},
\end{equation}
and its unilateral Laplace transform gives~\cite{englefieldSolutionSchrodingerEquation1968,baezSantosRevisitingLaplaceTransform2025}
\begin{equation}
  \left(\hbar s I + i H\right)\ket{\Psi(s)} = \hbar \ket{\psi(0)}.
\end{equation}
In this form, the reduction is obtained directly by Gaussian elimination, exactly as in classical network analysis~\cite{siebertCircuitsSignalsSystems1986,chenHaasElectricalInterpretationAlgorithm1962}.
This is the zero-input case of the more general inhomogeneous Schr\"odinger equation treated in Appendix~\ref{sec:app:inhomogeneous_norton}.
Time-dependent Hamiltonian blocks require a different reduction and are addressed in Appendix~\ref{sec:app:time_dependent}.

\subsection{\label{sec:qnt:gaussian} Single-Port Reduction by Gaussian Elimination}

Partitioning the state and Hamiltonian into a single retained coordinate \(P\) and an eliminated sector \(Q\) gives
\begin{equation}
  \begin{bmatrix}
    \hbar s + i H_{PP} & i H_{PQ}             \\
    i H_{QP}           & \hbar s I + i H_{QQ}
  \end{bmatrix}
  \begin{bmatrix}
    \Psi_P(s) \\
    \Psi_Q(s)
  \end{bmatrix}
  =
  \hbar
  \begin{bmatrix}
    \psi_P(0) \\
    \psi_Q(0)
  \end{bmatrix}.
\end{equation}
Gaussian elimination of the \(Q\)-sector yields the exact single-port reduction.

\begin{theorem}[The Quantum Norton Theorem]
  \label{thm:qnt_single_port}
  Let \(H\) be a finite-dimensional time-independent Hermitian Hamiltonian, and partition the Laplace domain Schr\"odinger equation so that one amplitude \(\Psi_P(s)\) is retained while all remaining amplitudes are grouped into an eliminated sector \(Q\).
  Then Gaussian elimination of the \(Q\)-sector yields
  \begin{equation}
    \left[\hbar s + i H_{PP} + \Sigma(s)\right]\Psi_P(s)
    =
    \hbar \psi_P(0) + F(s),
  \end{equation}
  where \(\Sigma(s)\) is the self-energy produced by the eliminated sector and \(F(s)\) is the corresponding reduced source term.
  Under the circuit mapping \(\Psi_P(s) \leftrightarrow V_P(s)\), \(\hbar \leftrightarrow C_\hbar\), \(i H_{PP} \leftrightarrow Y_{PP}\), \(\Sigma(s) \leftrightarrow Y_{\Sigma}(s)\), and \(F(s) \leftrightarrow I_F(s)\), this becomes
  \begin{equation}
    \left[C_\hbar s + Y_{PP} + Y_{\Sigma}(s)\right] V_P(s)
    =
    C_\hbar V_P(0) + I_F(s),
  \end{equation}
  where \(Y_{\Sigma}(s)\) is the Norton admittance associated with the self-energy \(\Sigma(s)\) and \(I_F(s)\) is the corresponding effective current source.
  Thus, the full \(Q\) sector may now be replaced at the retained port by the Norton pair \(\{Y_{\Sigma}(s),\,I_F(s)\}\).
\end{theorem}
So far, no approximation, perturbative expansion, or continuum limit has been applied.
In later sections we will see how such approximations can further simplify circuit implementations.

\subsection{\label{sec:qnt:tls} Two-Level System Reduction}

The simplest nontrivial instance of the theorem is a two-level system.
Consider again a two-level system with Hamiltonian
\begin{equation}
  H_\mathrm{TLS} =
  \begin{bmatrix}
    E_P & g   \\
    g^* & E_Q
  \end{bmatrix},
\end{equation}
where the first level is retained, and the second is eliminated.
In the Laplace domain, Gaussian elimination gives a single equation for \(\Psi_P(s)\) without explicit coupling to \(\Psi_Q(s)\):
\begin{equation}
  \begin{aligned}
     & \left(\hbar s + i E_P + \frac{\vab{g}^2}{\hbar s + i E_Q}\right) \Psi_P(s) \\
     & = \hbar \psi_P(0) - i\hbar \frac{g}{\hbar s + iE_Q} \psi_Q(0).
  \end{aligned}
\end{equation}
This is the reduced single-port equation
\begin{equation}
  \left[\hbar s + i E_P + \Sigma(s)\right] \Psi_P(s)
  = \hbar \psi_P(0) + F(s),
\end{equation}
with
\begin{equation}
  \Sigma(s) = \frac{\vab{g}^2}{\hbar s + i E_Q},
  \qquad
  F(s) = - i \hbar \frac{g}{\hbar s + iE_Q} \psi_Q(0).
\end{equation}

Thus, the eliminated level contributes a self-energy \(\Sigma(s)\) with a single pole at \(s=-iE_Q/\hbar\) and an effective source \(F(s)\) fixed by the eliminated initial amplitude.
At the retained port, the eliminated level is exactly equivalent to a Norton element with dynamical admittance \(Y_{\Sigma}(s)\) and current source \(I_F(s)\) in parallel.

\subsection{\label{sec:qnt:realization} Realization of Reduced Rational Terms}

\begin{figure}
  \centering
  \includegraphics[max width=\columnwidth]{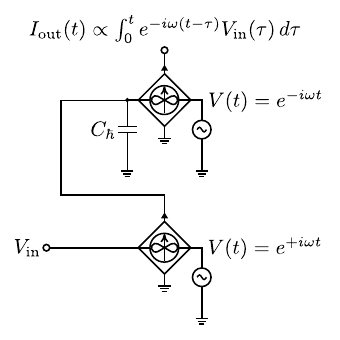}
  \caption{
    Circuit realization of a single-pole contribution to the reduced Norton admittance.
    For finite Hermitian systems, the single-port self-energy \(\Sigma(s)\) is a sum of simple pole contributions.
    For a circuit implementation, the dynamical admittance \(Y_{\Sigma}(s)\) may then be constructed from multiple subcircuits of this form connected in parallel, one for each nonzero pole contribution.
    The auxiliary node represents the internal dynamical mode associated with the pole, while the transadmittance mixers implement the weighted coupling to the retained coordinate.
  }
  \label{fig:single_pole_realization}
\end{figure}

\begin{figure}
  \centering
  \includegraphics[max width=\columnwidth]{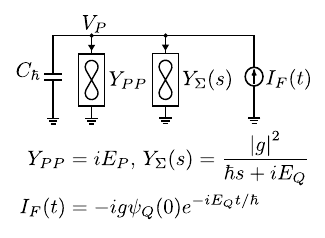}
  \caption{
    Circuit representation of the reduced two-level system (see Fig.~\ref{fig:tls_mapping} for the full circuit).
  }
  \label{fig:tls_reduced}
\end{figure}

The single-port equation derived from Gaussian elimination results in a self-energy term \(\Sigma(s)\) and a source term \(F(s)\).
For a finite eliminated sector, both terms are rational functions of \(s\), and an eliminated eigenvalue appears as a pole only if its projected weight is nonzero.
In the circuit implementation, \(\Sigma(s)\) is implemented as a reduced Norton admittance \(Y_{\Sigma}(s)\), while \(F(s)\) is implemented as a corresponding effective current source \(I_F(s)\).

The simplest case is a single-pole self-energy,
\begin{equation}
  \Sigma(s) = \frac{R}{\hbar s + i E},
\end{equation}
where \(R\geq0\) is the spectral weight set by the coupling to the eliminated sector.

In the time domain, this term acts as a causal memory kernel:
\begin{equation}
  \begin{aligned}
    \Sigma(s)\Psi_P(s)
     & \Rightarrow
    \int_0^t K_{\Sigma}(t-\tau)\psi_P(\tau)\,d\tau, \\
    K_{\Sigma}(t)
     & = \mathcal{L}^{-1}\{\Sigma(s)\}.
  \end{aligned}
\end{equation}
The \(Q\)-sector coordinates are effectively eliminated, but their influence on the \(P\)-sector remains in the form of a causal memory term and an effective source determined by \(\psi_Q(0)\).
For \(R>0\), one realization uses an auxiliary mode of energy \(E\) and conjugate couplings of magnitude \(\sqrt R\), as in Fig.~\ref{fig:single_pole_realization}.
Figure~\ref{fig:tls_reduced} shows the resulting reduced circuit for the two-level system.

Of course, combining the \(Q\) coordinates into an admittance and source does not necessarily reduce the complexity of the circuit implementation.
However, for cases with appropriate symmetry, or cases where approximations are applicable (e.g., sparse or weak coupling), the source and admittance components may then take advantage of those approximations.
\section{\label{sec:grover} Grover Search as a Single-Port Reduction}

We now apply the quantum Norton reduction to the continuous-time Grover search Hamiltonian~\cite{groverFastQuantumMechanical1996,farhiGutmannAnalogAnalogueDigital1998,fennerIntuitiveHamiltonianQuantum2000} with one marked state (the retained port) and three unmarked states that are eliminated via Gaussian elimination in the Laplace domain.
The reduction to a marked state coupled to a collective unmarked mode is well established~\cite{farhiGutmannAnalogAnalogueDigital1998,groverSenguptaClassicalAnalogQuantum2002}, and here we show how to realize this reduction as a Norton equivalent circuit with the unmarked sector represented as an effective load on the marked state.
The marked state is specified during circuit construction, so this is a demonstration of the reduced state dynamics and not a procedure for finding an unknown marked state.
The resulting single-port Norton equivalent circuit reproduces the exact marked state dynamics of the full Grover search circuit.

\subsection{\label{sec:grover:full} Four-State Grover Dynamics}

\begin{figure*}
  \centering
  \includegraphics[max width=\textwidth]{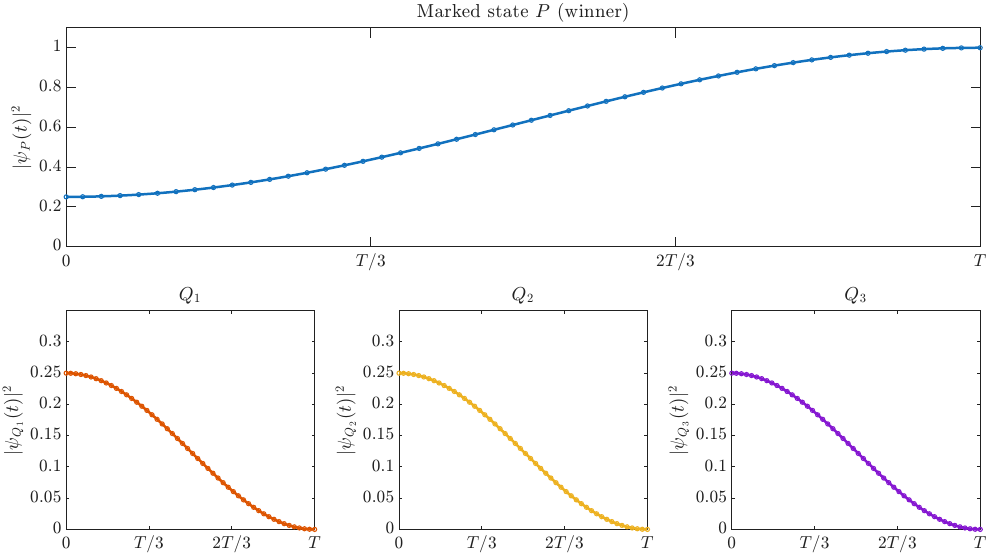}
  \caption{
    Full four-state Grover dynamics over the search window \(0 \le t \le T\), where \(T\) is the first maximum of the marked state probability.
    The solid curves show the exact Schr\"odinger evolution, while the circles show the circuit simulation of the full circuit realization.
    Over this interval, the marked state probability rises while the three unmarked state probabilities fall.
  }
  \label{fig:grover_full_plot}
\end{figure*}

\begin{figure}
  \centering
  \includegraphics[max width=\columnwidth]{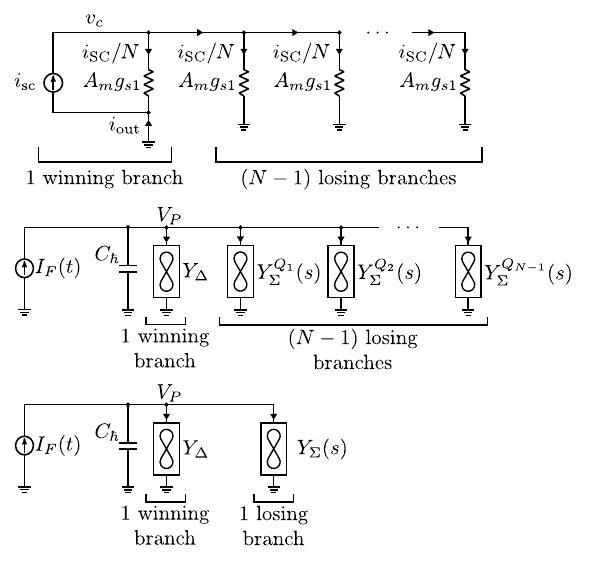}
  \caption{
    Winner-take-all interpretation of the four-state Grover problem over the search window.
    \emph{Top:} a classical winner-take-all Norton network with one distinguished branch and \((N-1)\) competing branches.
    \emph{Middle:} the unmarked sector's Norton admittance at the retained node \(P\), expressed as identical parallel contributions \(Y_{\Sigma}^{Q_1}(s),Y_{\Sigma}^{Q_2}(s),\dots,Y_{\Sigma}^{Q_{N-1}}(s)\).
    \emph{Bottom:} the unmarked sector's load contributions can be combined by symmetry into a single equivalent Norton admittance \(Y_{\Sigma}(s)\) and source \(I_F(s)\).
  }
  \label{fig:grover_wta}
\end{figure}

We consider a four-state search space with one marked state and three unmarked states.
We limit ourselves to four states here to illustrate the circuit construction, with arbitrary \(N\) handled in Sec.~\ref{sec:disordered_grover}.
We retain the marked basis state \(\ket*{\psi_P}\) as the retained port and write the ordered basis as
\begin{equation}
  \{\ket*{\psi_P},\ket*{\psi_{Q_1}},\ket*{\psi_{Q_2}},\ket*{\psi_{Q_3}}\}.
\end{equation}
In this basis, the Grover Hamiltonian is
\begin{equation}
  H =
  \begin{bmatrix}
    \Delta & \gamma & \gamma & \gamma \\
    \gamma & 0      & \gamma & \gamma \\
    \gamma & \gamma & 0      & \gamma \\
    \gamma & \gamma & \gamma & 0
  \end{bmatrix}.
\end{equation}
For this four-state example, we set \(\gamma=\Delta/4\) with diagonal entry \(\Delta\) for the retained state \(\ket*{\psi_P}\) and zero diagonal entries for the three equivalent unmarked states.

Starting from the uniform initial state, the marked sector draws amplitude from the unmarked sector over the search window \(0 \le t \le T\), where \(T\) is the first maximum of the marked state probability (for this choice of parameters, \(T = \pi\hbar/\Delta\)).
Figure~\ref{fig:grover_full_plot} shows that the circuit simulation reproduces the exact four-state Grover dynamics over the full search window.

From the viewpoint of the retained port, this is a winner-take-all configuration where the predesignated marked state is the winner and the three unmarked states form a symmetric set of competing branches, as illustrated in the top panel of Fig.~\ref{fig:grover_wta}.
The middle panel shows the unmarked sector's Norton admittances as three parallel admittance contributions, obtained through Gaussian elimination.
By symmetry, the three unmarked branches combine to a single equivalent admittance \(Y_{\Sigma}(s)\) as shown in the bottom panel.

\subsection{\label{sec:grover:reduced} Exact Reduction of the Unmarked Sector}

\begin{figure}
  \centering
  \includegraphics[max width=\columnwidth]{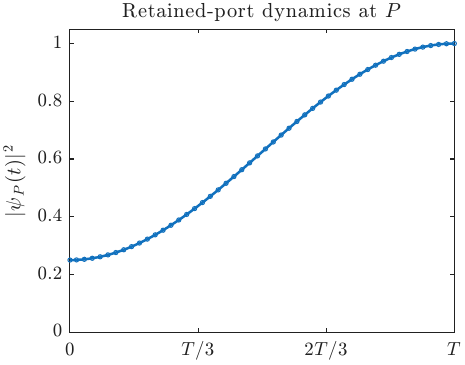}
  \caption{
    Marked state dynamics for the full and reduced Grover models over the search window \(0 \le t \le T\).
    The solid curve shows the exact probability \(\vab{\psi_P(t)}^2\), while the circles show the circuit simulation of the reduced Norton circuit.
    The reduced circuit reproduces these dynamics exactly.
  }
  \label{fig:grover_reduced_plot}
\end{figure}

Starting from the Laplace domain Schr\"odinger equation,
\begin{equation}
  \left(\hbar s I + iH\right)\ket*{\Psi(s)} = \hbar \ket*{\psi(0)},
\end{equation}
Gaussian elimination of the three \(Q\)-sector equations gives an exact single-port equation for the retained amplitude \(\Psi_P(s)\):
\begin{equation}
  \begin{aligned}
     & \left[\hbar s + i\Delta + \frac{3\gamma^2}{\hbar s + 2i\gamma}\right]\Psi_P(s) \\
     & = \hbar \psi_P(0) - \frac{i\hbar\gamma}{\hbar s + 2i\gamma}
    \left[\psi_{Q_1}(0)+\psi_{Q_2}(0)+\psi_{Q_3}(0)\right].
  \end{aligned}
\end{equation}

The resulting self-energy is then the single expression
\begin{equation}
  \Sigma(s) = \frac{3\gamma^2}{\hbar s + 2i\gamma}.
\end{equation}
The self-energy appears in the Norton equivalent circuit as a single admittance \(Y_{\Sigma}(s)\) shown in Fig.~\ref{fig:grover_wta}.

The corresponding reduced source term is
\begin{equation}
  \begin{aligned}
    F(s) & = - \frac{i\hbar\gamma}{\hbar s + 2i\gamma} \left[\psi_{Q_1}(0)+\psi_{Q_2}(0)+\psi_{Q_3}(0)\right] \\
         & = \frac{-3i\hbar\gamma/2}{\hbar s + 2i\gamma}
  \end{aligned}
\end{equation}
for the uniform initial state. In the circuit realization, this source term appears as the Norton current source \(I_F(s)\).
Figure~\ref{fig:grover_reduced_plot} compares the marked state dynamics of the reduced circuit with those of the exact four-state model.
\section{\label{sec:general_qnt} Generalized Quantum Norton Theorem}

\begin{figure*}
  \centering
  \includegraphics[max width=\textwidth]{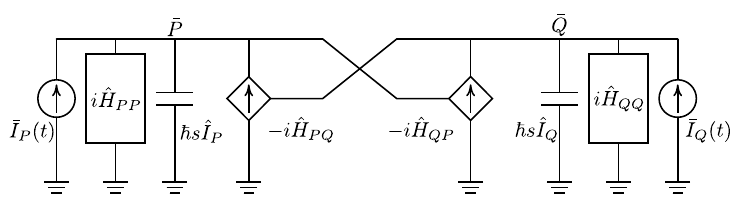}
  \caption{
    The figure shows a vector-matrix complex circuit representation that provides circuit and physical intuition for our mathematical equations.
    In particular, the Norton theorem can be viewed as a vector gyration or reflected admittance of the eliminated \(Q\) sector onto \(P\) for the Norton admittance, and a similar reflection of driven and initial condition currents at \(\bar{Q}\) onto \(\bar{P}\).
    In particular, we have \(\bar{I}_Q(t) = \bar{I}^\delta_Q(t) + \bar{I}^D_Q(t)\), \(\int_{-\infty}^{+\infty} \bar{I}^\delta_Q(t) \, dt = \hbar \bar{Q}(0)\), \(\bar{I}_P(t) = \bar{I}^\delta_P(t) + \bar{I}^D_P(t)\), and \(\int_{-\infty}^{+\infty} \bar{I}^\delta_P(t) \, dt = \hbar \bar{P}(0)\).
  }
  \label{fig:gyrations}
\end{figure*}

The single-port quantum Norton theorem can be generalized to an arbitrary number of retained coordinates using the Schur complement in Laplace space~\cite{golubVanLoanMatrixComputations2013}.
In classical network theory, the same construction is known as Kron reduction~\cite{kronTensorAnalysisNetworks1939,dorflerBulloKronReductionGraphs2013}.

Partitioning the state and Hamiltonian into retained (\(P\)) and eliminated (\(Q\)) sectors puts the Laplace domain Schr\"odinger equation in the form
\begin{equation}
  \begin{bmatrix}
    \hbar s I + i H_{PP} & i H_{PQ}             \\
    i H_{QP}             & \hbar s I + i H_{QQ}
  \end{bmatrix}
  \begin{bmatrix}
    \Psi_P(s) \\
    \Psi_Q(s)
  \end{bmatrix}
  =
  \hbar
  \begin{bmatrix}
    \psi_P(0) \\
    \psi_Q(0)
  \end{bmatrix}.
\end{equation}
Taking the Schur complement of the \(Q\) block gives the following result.

\begin{theorem}[Generalized Quantum Norton Theorem]
  \label{thm:qnt_multiport}
  Let \(H\) be a finite-dimensional, time-independent Hermitian Hamiltonian acting on a Hilbert space partitioned into orthogonal retained subspace \(P\) and eliminated subspace \(Q\).
  Then the exact reduced dynamics on the retained subspace are
  \begin{equation}
    \left[\hbar s I + i H_{PP} + \Sigma(s)\right] \ket{\Psi_P(s)} = \hbar \ket{\psi_P(0)} + F(s),
  \end{equation}
  where
  \begin{equation}
    \begin{aligned}
      \Sigma(s) & = H_{PQ} \left(\hbar s I + i H_{QQ}\right)^{-1} H_{QP},                   \\
      F(s)      & = -i \hbar H_{PQ} \left(\hbar s I + i H_{QQ}\right)^{-1} \ket{\psi_Q(0)}.
    \end{aligned}
    \label{eq:generalized_norton}
  \end{equation}
  At the retained ports, the eliminated sector is exactly represented by a Norton multiport circuit with admittance matrix \(Y_{\Sigma}(s)\) corresponding to \(\Sigma(s)\) and current source vector \(I_F(s)\) corresponding to \(F(s)\).
\end{theorem}

Figure~\ref{fig:gyrations} has an interesting interpretation of Eq.~\ref{eq:generalized_norton} that is intuitive and useful from a circuits perspective, using the well-known concept of gyration.
Typically, gyration is used to create inductors from capacitors and transconductors; or to transform/reflect a terminating load  impedance from a short at one end of a quarter-wave transmission line to an open or vice versa; or to have the load impedance in the secondary of a resonant near-field RF link reflect as an effective impedance in the primary~\cite{sarpeshkarUltraLowPower2010}.
The fundamental pivot-based operations of Gaussian elimination, have multiple gyratory terms of the form (\(\text{column}/\text{pivot} \times \text{row}\)) that reflect terms of a pivot row into rows below, to zero out, one by one, all columns of the latter row.
The net effect is to make a matrix upper triangular and remove all of its feedback loops.
It is then easy to obtain solutions to matrix equations.
If, in Gaussian elimination, we map scalar variables to complex vector variables, multiplication operations to vector matrix multiplication, the pivot to a net quantum load admittance, and division by the pivot to an inverse, then Eq.~\ref{eq:generalized_norton} is seen to just be a complex gyratory reflection of the eliminated sector \(Q\) onto \(P\).
The Norton admittance is the reflection of admittances at \(Q\) into \(P\), while the Norton source is the reflection of initial conditions and sources at \(Q\) onto \(P\).
Kirchhoff's current law on such vector variables corresponds to addition.
Thus, the content of the generalized Norton theorem has a simple vector circuit interpretation.
As we show in a later section, the effect of environmental noise in environmental vector admittances in the eliminated sector (which can be lossy and therefore fluctuating), then reflects into an effective vector admittance (the Norton admittance) in the retained sector.
Successive reflections (like successive eliminations in Gaussian elimination) create the net effective Norton admittance and net effective Norton source.
Both initial conditions and driving terms port correctly as the caption in Fig.~\ref{fig:gyrations} shows.
For simplicity, in Eq.~\ref{eq:generalized_norton}, we have only shown how the initial condition at \(Q\) ports to \(P\); Appendix~\ref{sec:app:inhomogeneous_norton}, Eq.~\ref{eq:norton_driven} adds a term equal to \(H_{PQ} \left( \hbar s I + i H_{QQ} \right)^{-1} \eta_Q(s)\), which shows how the driving source at \(Q\) ports to \(P\).
If the latter term is added to the initial-condition term, we have the complete Norton theorem for driving terms and initial conditions.

Note that all vectors and matrices in Fig.~\ref{fig:gyrations} have components with complex numbers such that we have quantum dynamics, not classical dynamics.
The classical analog instantiation of this circuit would need paired real and imaginary variables as in Fig.~\ref{fig:overview} and corresponding \(2N \times 2N\) maps of the \(N \times N\) complex matrices.

To compare with the standard Green's function formulation, write the retained block of \(G(z)=(zI-H)^{-1}\) as
\begin{equation}
  G_{PP}(z)= \left[z I-H_{PP} - H_{PQ} (zI-H_{QQ})^{-1} H_{QP}\right]^{-1}
\end{equation}
wherever the required inverses exist.
The standard embedding self-energy \(H_{PQ} (z I - H_{QQ})^{-1 } H_{QP}\) of Green's function methods~\cite{economouGreensFunctionsQuantum2006,dattaQuantumTransportAtom2005} equals \(-i \Sigma(s)\) under the change of variable \(z = i \hbar s\).

Let \(\ket*{\Psi_Q^{(0)}(s)}\) denote the Laplace transform of the \(Q\)-sector state evolving under \(H_{QQ}\) alone, with initial condition \(\ket{\psi_Q(0)}\).
Then
\begin{equation}
  \left(\hbar s I + i H_{QQ}\right) \ket*{\Psi_Q^{(0)}(s)} = \hbar \ket{\psi_Q(0)},
\end{equation}
so that
\begin{equation}
  \ket*{\Psi_Q^{(0)}(s)} = \hbar \left(\hbar s I + i H_{QQ}\right)^{-1} \ket{\psi_Q(0)}.
\end{equation}
Thus, \(\left(\hbar s I + i H_{QQ}\right)^{-1}\) is the resolvent of the isolated eliminated sector, so that the reduced source can be written as
\begin{equation}
  F(s) = -i H_{PQ} \ket*{\Psi_Q^{(0)}(s)}.
\end{equation}

Because \(H_{QQ}\) is finite-dimensional and Hermitian, its spectral decomposition is
\begin{equation}
  H_{QQ} = \sum_k E_k \Pi_k,
\end{equation}
where \(E_k\) are the distinct real eigenvalues of \(H_{QQ}\) and \(\Pi_k\) are the corresponding orthogonal spectral projectors.
Hence, for \(s \neq -iE_k/\hbar\),
\begin{equation}
  \left(\hbar s I + i H_{QQ}\right)^{-1} = \sum_k \frac{1}{\hbar s + i E_k} \Pi_k.
\end{equation}

Substituting this into the definitions of \(\Sigma(s)\) and \(F(s)\) gives
\begin{equation}
  \begin{aligned}
    \Sigma(s) & = \sum_k \frac{1}{\hbar s + i E_k} H_{PQ} \Pi_k H_{QP},                  \\
    F(s)      & = -i \hbar H_{PQ} \sum_k \frac{1}{\hbar s + iE_k} \Pi_k \ket{\psi_Q(0)}.
  \end{aligned}
\end{equation}
The corresponding residue matrix in the self-energy expansion is
\begin{equation}
  R_k = H_{PQ} \Pi_k H_{QP},
\end{equation}
which acts on the retained subspace and reduces to a scalar in the single-port case.
Both the Norton admittance and current source inherit their poles from the spectrum of the eliminated sector.
\section{\label{sec:pole_structure} Pole Structure of Reduced Quantum Dynamics}

The generalized Norton theorem of Sec.~\ref{sec:general_qnt} reduces the retained dynamics to the pole-residue structure of the reduced self-energy \(\Sigma(s)\) and source term \(F(s)\).
The spectrum of \(H_{QQ}\) fixes the pole locations, while the projected couplings \(H_{PQ}\Pi_k H_{QP}\) and \(H_{PQ}\Pi_k\ket{\psi_Q(0)}\) fix the reduced self-energy and source weights.
In circuit terms, this pole-residue structure fixes the reduced Norton realization: each pole contributes a dynamical branch, and the associated residue determines its coupling to the retained coordinates.

The examples below specialize this general result to single-port reductions of increasing spectral complexity.

\subsection{\label{sec:pole_structure:tls} Single Pole --- Two-Level System}

\begin{figure}
  \centering
  \includegraphics[max width=\columnwidth]{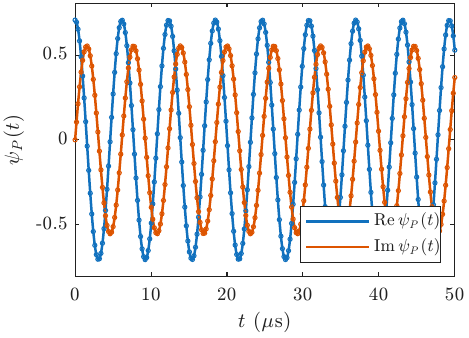}
  \caption{
    Reduced-circuit dynamics of the retained amplitude \(\psi_P(t)\), showing exact agreement with the retained-coordinate dynamics of the full quantum system (solid lines).
  }
  \label{fig:tls_reduced:plot}
\end{figure}

The two-level system of Sec.~\ref{sec:qnt:tls} provides the minimal one-pole example.
Its reduced self-energy and source term are
\begin{equation}
  \begin{aligned}
    \Sigma(s) & = \frac{\vab{g}^2}{\hbar s + iE_Q},          \\
    F(s)      & = -i\hbar \frac{g}{\hbar s + iE_Q}\psi_Q(0).
  \end{aligned}
\end{equation}
Both are controlled by the same pole at \(s=-iE_Q/\hbar\).
Figure~\ref{fig:tls_reduced:plot} shows that this one-pole reduction exactly reproduces the retained-coordinate dynamics of the full system.

\subsection{\label{sec:pole_structure:static} Static Approximation}

\begin{figure}
  \centering
  \includegraphics[max width=\columnwidth]{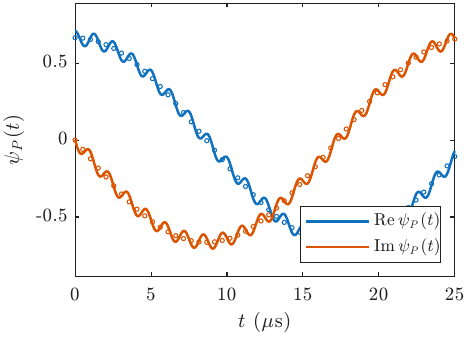}
  \caption{
    Reduced dynamics obtained from the static approximation to both \(\Sigma(s)\) and \(F(s)\), compared with the exact full two-level dynamics (solid lines).
    The approximation reproduces the overall retained-coordinate dynamics when the eliminated mode lies far outside the frequency range relevant to the retained dynamics.
  }
  \label{fig:tls_static:plot}
\end{figure}

Now consider the regime \(\vab{E_P/\hbar} \ll \vab{E_Q/\hbar}\), where the eliminated mode evolves much faster than the retained one.
Then the pole of \(\Sigma(s)\) lies far outside the frequency range relevant to the retained dynamics.
The reduced self-energy and source term therefore admit low-frequency expansions,
\begin{equation}
  \begin{aligned}
    \Sigma(s) & = \frac{\vab{g}^2}{\hbar s + iE_Q}                                                              \\
              & = -\frac{i\vab{g}^2}{E_Q} + \mathcal{O}\!\left(\frac{\vab{g}^2 \hbar s}{E_Q^2}\right),          \\
    F(s)      & = -i\hbar \frac{g}{\hbar s + iE_Q}\psi_Q(0)                                                     \\
              & = -\frac{\hbar g}{E_Q}\psi_Q(0) + \mathcal{O}\!\left(\frac{\hbar^2 g s}{E_Q^2}\psi_Q(0)\right).
  \end{aligned}
\end{equation}
To leading order,
\begin{equation}
  \Sigma(s) \approx -\frac{i\vab{g}^2}{E_Q},
  \qquad
  F(s) \approx -\frac{\hbar g}{E_Q}\psi_Q(0).
\end{equation}
The leading-order source term may be absorbed into an effective retained initial condition,
\begin{equation}
  \psi_{P,\mathrm{eff}}(0) = \psi_P(0) - \frac{g}{E_Q}\psi_Q(0).
\end{equation}

In this limit, the exact one-pole Norton reduction reduces to a static approximation~\cite{lowdinStudiesPerturbationTheory1962,bravyiLossSchriefferWolffTransformation2011}.
To leading order, \(\Sigma(s)\) renormalizes the retained energy, while \(F(s)\) becomes \(s\)-independent and is absorbed into an effective retained initial condition.
In circuit terms, this replaces the reduced load by a constant admittance and renormalizes the retained initial condition.
Figure~\ref{fig:tls_static:plot} shows that this approximation reproduces the overall retained-coordinate dynamics of the full system.

\subsection{\label{sec:pole_structure:chain} Two Poles --- Short Chain}

\begin{figure}
  \centering
  \includegraphics[max width=\columnwidth]{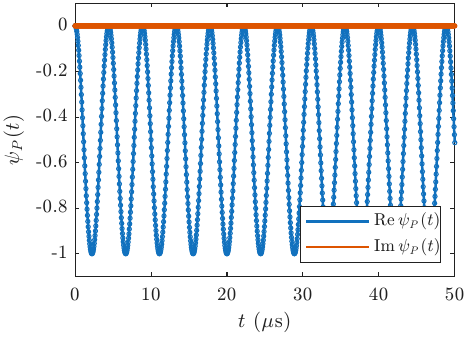}
  \caption{
    Reduced two-pole dynamics of the retained amplitude \(\psi_P(t)\), showing exact agreement with the full finite-chain dynamics (solid lines).
  }
  \label{fig:chain:plot}
\end{figure}

Next consider a three-site nearest-neighbor tight-binding chain~\cite{economouGreensFunctionsQuantum2006},
\begin{equation}
  H =
  \begin{bmatrix}
    0 & J & 0 \\
    J & 0 & J \\
    0 & J & 0
  \end{bmatrix}
  .
\end{equation}
Retaining the first site and eliminating the last two gives
\begin{equation}
  \Sigma(s) = \frac{\hbar J^2 s}{\hbar^2 s^2 + J^2}
  = \frac{J^2 / 2}{\hbar s + i J} + \frac{J^2 / 2}{\hbar s - i J},
\end{equation}
with poles at \(s=\pm iJ/\hbar\) and equal residues \(J^2/2\).

The reduced operator contains two poles because the eliminated block \(H_{QQ}\) has two eigenvalues.
Equivalently, the pole expansion has two projector contributions \(H_{PQ}\Pi_k H_{QP}\), one for each eliminated eigenmode.
The equal residues show that both modes couple equally to the retained coordinate, reflecting the symmetry of the chain and the choice of retained boundary site.
For the initial condition used here, the reduced source term vanishes, so the exact reduced response is fixed entirely by these two spectral branches.
Figure~\ref{fig:chain:plot} shows that the resulting two-pole reduction exactly reproduces the retained-coordinate dynamics of the full chain.
Appendix~\ref{sec:app:finite_chain} gives the pole expansion for a finite chain of arbitrary length.
Below, we consider the limit as the chain length tends to infinity.

\subsection{\label{sec:pole_structure:semiinfinite} Many Poles --- Semi-Infinite Chain}

\begin{figure}
  \centering
  \includegraphics[max width=\columnwidth]{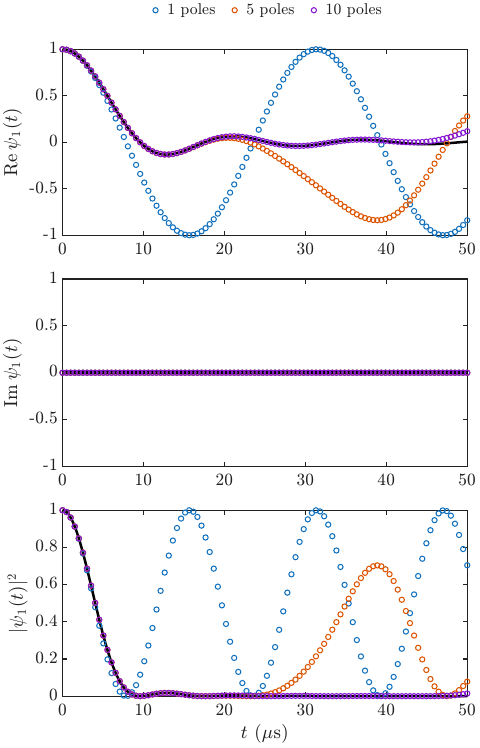}
  \caption{
    Boundary dynamics of the semi-infinite chain obtained from finite-pole reduced models, compared with the exact full-system solution (solid black line).
    As the number of poles increases, the reduced dynamics converge to the exact boundary dynamics.
  }
  \label{fig:semi_infinite_chain:plot}
\end{figure}

As an infinite-length extension of the finite-dimensional construction above, consider a semi-infinite tight-binding chain with the first coordinate retained and the remainder eliminated.
Eliminating the first site of the remainder again exposes a semi-infinite chain with the same nearest-neighbor structure.
The self-energy therefore satisfies the self-consistency relation
\begin{equation}
  \Sigma(s) = H_{PQ}\left(\hbar s I + iH_{QQ}\right)^{-1}H_{QP} = \frac{J^2}{\hbar s + \Sigma(s)}.
\end{equation}
Solving the resulting quadratic equation gives
\begin{equation}
  \Sigma(s) = \frac{1}{2}\left(-\hbar s + \sqrt{\hbar^2 s^2 + 4J^2}\right),
\end{equation}
where the physical branch is fixed by the high-frequency condition \(\Sigma(s)\to 0\) as \(\vab{s}\to\infty\), equivalently \(\Sigma(s)\sim J^2/(\hbar s)\).
For the boundary-localized initial condition used here, the reduced source term vanishes, so the reduced dynamics are determined entirely by \(\Sigma(s)\).

This self-similar construction is the semi-infinite-chain analog of the chain mappings used in open quantum systems, where structured reservoirs are represented as semi-infinite chains via orthogonal polynomials~\cite{chinPlenioExactMappingSystemreservoir2010}.
It is the infinite-length limit of the finite-chain pole expansion
\begin{equation}
  \Sigma_L(s) = \sum_{k=1}^{L} \frac{R_k}{\hbar s + iE_k}.
\end{equation}
For a chain of length \(L\), the eigenvalues \(E_k = 2J\cos(\pi k/(L+1))\) sample the interval \([-2J,2J]\) (Appendix~\ref{sec:app:finite_chain} and Appendix~\ref{sec:app:infinite_chain}).
As \(L\to\infty\), these poles become dense and the discrete sum approaches a continuum.
In the complex \(s\)-plane, the discrete pole set condenses into a branch cut associated with the continuous band of the eliminated subsystem.

In the time domain, the boundary response admits a closed form in terms of Bessel functions, as is well known in continuous-time quantum walk models~\cite{ben-avrahamTamonOnedimensionalContinuoustimeQuantum2004,economouGreensFunctionsQuantum2006}.
The circuit realization approximates this continuum by truncating the pole expansion to a finite number of modes.
Figure~\ref{fig:semi_infinite_chain:plot} shows the convergence of that approximation to the exact continuous response.
\section{\label{sec:disordered_grover} Disordered Grover Search: Bright and Dark Sectors}

\begin{figure}
  \centering
  \includegraphics[max width=\columnwidth]{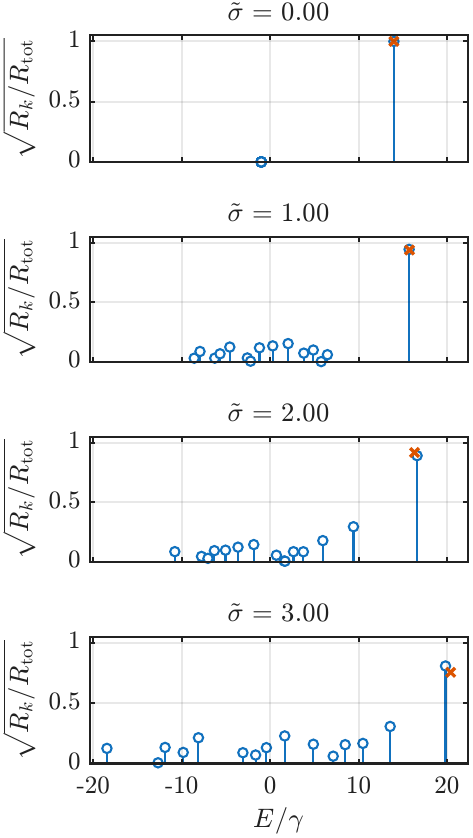}
  \caption{
    Pole leakage in the disordered Grover \(Q\)-sector for \(N=16\).
    The blue stems and open circles show the exact \(Q\)-sector pole energies and normalized residues, plotted as \(\sqrt{R_k/R_{\mathrm{tot}}}\) versus \(E/\gamma\), where \(R_{\mathrm{tot}}=(N-1)\gamma^2\) (the square root scale makes the small dark sector residues more distinguishable).
    The orange \(\times\) marks the bright-pole residue predicted by standard perturbation theory.
    From top to bottom, the panels are arranged in increasing disorder strength \(\tilde{\sigma}\).
    In the ideal limit (\(\tilde{\sigma}=0\)), all coupling weight at the boundary is concentrated in a single bright pole, while the dark sector carries zero residue.
    As \(\tilde{\sigma}\) increases, residue is redistributed into the dark sector and the bright-pole residue is reduced.
  }
  \label{fig:disordered_grover_poles_plot}
\end{figure}

\begin{figure}
  \centering
  \includegraphics[max width=\columnwidth]{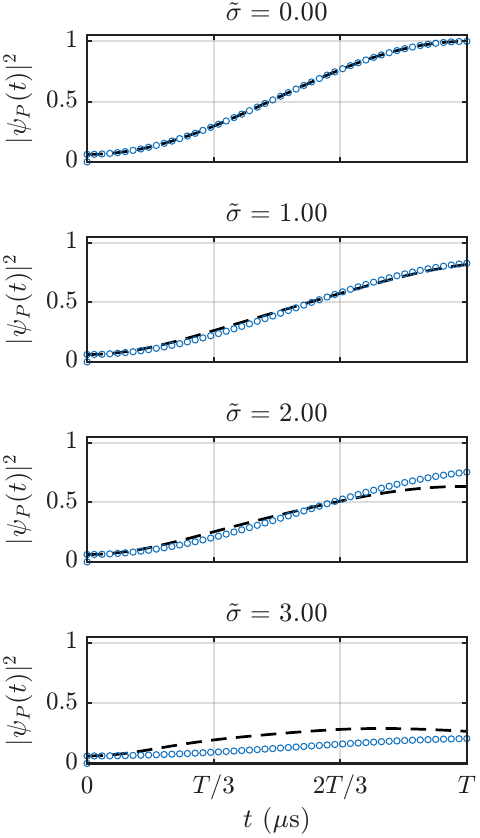}
  \caption{
    Search probability for the disordered Grover network with \(N=16\).
    The black dashed curves show the exact marked-state probability \(|\psi_P(t)|^2\) from the full centered-disorder Hamiltonian, while the blue markers show the bright-only reduced model implemented in the circuit simulation.
    The horizontal axis is shown over the ideal search window \(0 \le t \le T\), where \(T\) is the ideal measurement time.
    From top to bottom, the panels are arranged in increasing disorder strength \(\tilde{\sigma}\).
    For weak disorder, the bright-only circuit closely tracks the full model's marked state probability.
    At larger \(\tilde{\sigma}\), the bright-only approximation still roughly matches the exact dynamics, but the exact dynamics now strongly differ from the ideal case.
  }
  \label{fig:disordered_grover_transient}
\end{figure}

Section~\ref{sec:grover} established the exact single-port reduction of Grover dynamics by retaining the marked coordinate and eliminating the unmarked sector.
Here we determine how diagonal disorder modifies that reduced picture~\cite{shenviWhaleyEffectsRandomNoisy2003,reitznerHilleryGroverSearchLocalized2019}.
In the ideal case, the eliminated sector contributes a single collective bright mode, while the orthogonal dark sector is uncoupled to the retained port.
Disorder mixes the bright and dark components, transferring residue out of the bright pole so that the single-pole description becomes an approximation.

For an \(N\)-state Grover problem, we retain the marked basis state \(\ket{\psi_P}\) and eliminate the unmarked sector \(Q\).
With coupling choice \(\gamma=\Delta/N\), the clean \(H_{QQ}\) Hamiltonian is
\begin{equation}
  H_{QQ} = \gamma(J_{N-1}-I_{N-1}),
\end{equation}
where \(J_{N-1}\) is the \((N-1)\times(N-1)\) all-ones matrix.
The coupling to the retained coordinate is
\begin{equation}
  H_{PQ} = \gamma[1,\dots,1] = \gamma\sqrt{N-1}\bra{\psi_r},
\end{equation}
where
\begin{equation}
  \ket{\psi_r}=\frac{1}{\sqrt{N-1}}\sum_{j}\ket{\psi_{Q_j}}
\end{equation}
is the symmetric unmarked state.

This structure splits the unmarked sector into a single bright mode and a dark manifold: since \(H_{PQ}=\gamma\sqrt{N-1}\bra{\psi_r}\), only the symmetric unmarked state \(\ket{\psi_r}\) couples to the retained coordinate, and its orthogonal complement is dark.
In the ideal case, \(\ket{\psi_r}\) is the unique bright eigenmode of \(H_{QQ}\) with eigenvalue \((N-2)\gamma\), while every dark state has eigenvalue \(-\gamma\) but zero coupling to the retained coordinate.
Further details are given in Appendix~\ref{sec:app:grover_bright}.

For the uniform initial state
\begin{equation}
  \begin{aligned}
    \ket{\psi(0)} & = \frac{1}{\sqrt{N}}\left(\ket{\psi_P}+\sum_{j}\ket{\psi_{Q_j}}\right)  \\
                  & = \frac{1}{\sqrt{N}}\left(\ket{\psi_P}+\sqrt{N-1}\ket{\psi_{r}}\right),
  \end{aligned}
\end{equation}
the clean reduced quantities are
\begin{equation}
  \begin{aligned}
    \Sigma_{\mathrm{clean}}(s)
     & = \frac{(N-1)\gamma^2}{\hbar s+i(N-2)\gamma}, \\
    F_{\mathrm{clean}}(s)
     & = -i\hbar \frac{(N-1)\gamma}{\sqrt{N}}
    \frac{1}{\hbar s+i(N-2)\gamma}.
  \end{aligned}
\end{equation}

We now add diagonal disorder on the unmarked states,
\begin{equation}
  H \to H+\sum_{j=1}^{N-1}\delta_{Q_j}\proj{\psi_{Q_j}}
\end{equation}
and impose the centering condition
\begin{equation}
  \sum_{j=1}^{N-1}\delta_{Q_j}=0.
\end{equation}
This removes the common-mode shift and isolates the effect of bright-dark mixing.
We also define
\begin{equation}
  \begin{aligned}
    \sigma^2
     & = \frac{1}{N-1}\sum_{j=1}^{N-1}\delta_{Q_j}^2, \\
    \tilde{\sigma}
     & = \frac{\sigma}{\sqrt{N-1}\gamma}.
  \end{aligned}
\end{equation}

Keeping the same uniform initial state, the reduced quantities under disorder take the pole-expanded form
\begin{equation}
  \begin{aligned}
    \Sigma(s)
     & = \sum_{k=1}^{N-1}\frac{R_k}{\hbar s+iE_k}, \\
    F(s)
     & = -i\hbar \frac{1}{\gamma \sqrt{N}}
    \sum_{k=1}^{N-1}\frac{R_k}{\hbar s+iE_k},
  \end{aligned}
\end{equation}
where \(E_k\) are the exact \(Q\)-sector pole energies and \(R_k\) are the corresponding residues.

Figure~\ref{fig:disordered_grover_poles_plot} shows this redistribution for representative centered disorder realizations at \(N=16\).
At \(\tilde{\sigma}=0\), the dark sector is present but uncoupled to the retained port.
As \(\tilde{\sigma}\) increases, the dark poles spread, acquire nonzero residues, and deplete the bright-pole residue.
The orange marker shows the perturbative bright-pole approximation derived in Appendix~\ref{sec:app:grover_bright}.

Figure~\ref{fig:disordered_grover_transient} shows the corresponding time-domain effect of dark sector leakage on the bright-only reduction.
For weak disorder, the renormalized bright pole reproduces the marked-state probability accurately.
At larger \(\tilde{\sigma}\), the bright-only model still captures the broad trend of the exact dynamics.
However, disorder suppresses the peak marked state probability over the ideal search window.
\section{\label{sec:conditioning} Reduced Dynamics of Composite Quantum Systems}

For a composite system, we choose an orthonormal environment basis and apply the Norton reduction to the system block associated with each environment basis state.
For pure states we eliminate the amplitudes outside that block, and for density matrix states we move to Liouville space and eliminate the other environment blocks as well as the coherences between blocks.
The resulting system density matrix contributions sum to the reduced state \(\rho_S\), giving the partial trace over the environment~\cite{breuerPetruccioneTheoryOpenQuantum2010,nielsenChuangQuantumComputationQuantum2010}.

\subsection{\label{sec:composite:tls_pure} Coupled Two-Level Systems: Pure State Example}

\begin{figure}
  \centering
  \includegraphics[max width=\columnwidth]{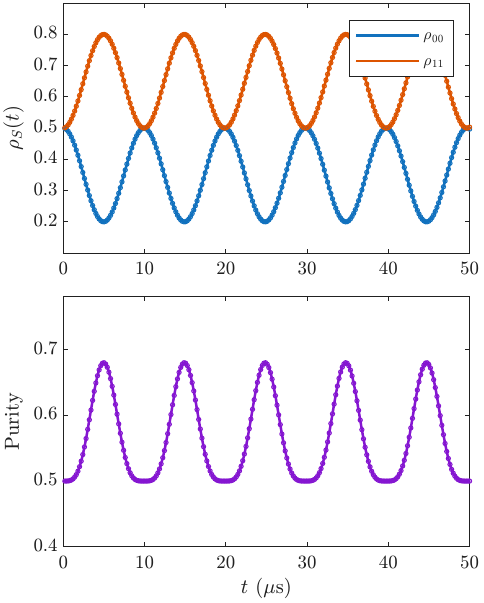}
  \caption{
    Reduced dynamics of a system TLS coupled to an environment TLS for an initial pure joint state.
    Shown are the system populations and purity \(\Tr\left(\rho_S^2\right)\) obtained from the full unitary Schr\"odinger evolution of the joint pure state (solid lines).
    Here the purity quantifies the mixedness of the reduced system state induced by entanglement with the environment.
  }
  \label{fig:tls_tls_pure}
\end{figure}

We begin with the simplest closed composite system: two coupled two-level systems (TLS), one designated as the system (\(S\)) and the other as the environment (\(E\)).
We work in the tensor-product basis
\begin{equation}
  \left\{\ket{0_E 0_S}, \ket{0_E 1_S}, \ket{1_E 0_S}, \ket{1_E 1_S}\right\},
\end{equation}
where the first index corresponds to the environment and the second to the system.
The Hamiltonian is
\begin{equation}
  H =
  \begin{bmatrix}
    0 & 0   & 0   & 0         \\
    0 & E_S & g   & 0         \\
    0 & g   & E_E & 0         \\
    0 & 0   & 0   & E_E + E_S
  \end{bmatrix},
\end{equation}
so that the interaction mixes the single-excitation states \(\ket{0_E 1_S}\) and \(\ket{1_E 0_S}\).

For a pure joint state
\begin{equation}
  \ket{\psi(t)} = \sum_e \sum_s c_{es}(t)\ket{e,s},
\end{equation}
we condition on each environment value \(e\) by defining the corresponding unnormalized system ket
\begin{equation}
  \ket*{\psi_S^{(e)}(t)} = \sum_s c_{es}(t)\ket{s}.
\end{equation}
From this contribution we construct
\begin{equation}
  \rho_S^{(e)}(t) = \ketbra*{\psi_S^{(e)}(t)}{\psi_S^{(e)}(t)} = \sum_{s,s'} c_{es}(t)c_{es'}^*(t)\ketbra{s}{s'},
\end{equation}
and the reduced state is obtained by summing over the retained environment values,
\begin{equation}
  \rho_S(t) = \sum_e \rho_S^{(e)}(t).
\end{equation}
For an orthonormal environment basis this is simply the partial trace,
\begin{equation}
  \rho_S(t) = \sum_e \braket[3]{e}{\rho(t)}{e}.
\end{equation}
The fixed environment value decomposition depends on the chosen basis, but the summed result is exactly the basis-independent reduced density matrix.
This reconstruction from separate environment-resolved contributions differs from the standard product-state projection used in the Nakajima-Zwanzig formalism, \(\mathcal P X=\rho_E^{\mathrm{ref}}\otimes\operatorname{Tr}_E X\), which maps the joint density operator to \(\rho_E^{\mathrm{ref}}\otimes\rho_S\), with \(\rho_E^{\mathrm{ref}}\) a fixed, normalized environment state~\cite{smirneVacchiniNakajimazwanzigTimeconvolutionlessMaster2010}.

For the circuit realization, we do not evolve the full composite state directly.
Instead, for each retained environment value \(e\), we solve the corresponding reduced equation with self-energy \(\Sigma^{(e)}(s)\) and source term \(F^{(e)}(s)\), and then assemble the reduced state from the resulting fixed environment contributions.
For this two-TLS Hamiltonian, the only nontrivial coupling is \(\ket{0_E 1_S}\leftrightarrow\ket{1_E 0_S}\), so the reduced self-energies take the explicit form
\begin{equation}
  \begin{aligned}
    \Sigma^{(e=0)}(s) & = \frac{g^2}{\hbar s + iE_E}\proj{0_E 1_S}, \\
    \Sigma^{(e=1)}(s) & = \frac{g^2}{\hbar s + iE_S}\proj{1_E 0_S}.
  \end{aligned}
\end{equation}
The corresponding source terms \(F^{(e=0)}(s)\) and \(F^{(e=1)}(s)\) encode the initial amplitudes in the eliminated sectors.
Figure~\ref{fig:tls_tls_pure} shows that the reconstructed system populations and reduced-state purity \(\Tr(\rho_S^2)\) agree with those obtained from exact Schr\"odinger evolution of the full composite pure state.

\subsection{\label{sec:composite:mixed} Mixed State Extension and Liouville Space Formulation}

\begin{figure}
  \centering
  \includegraphics[max width=\columnwidth]{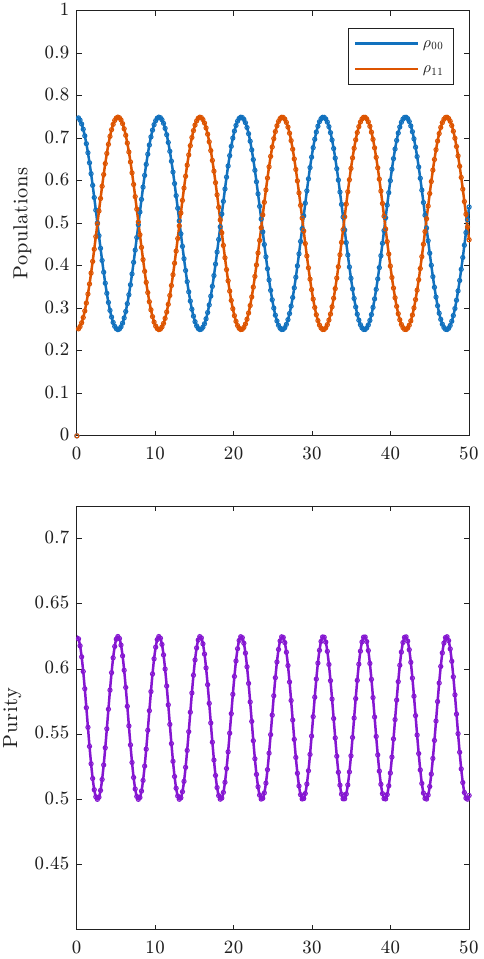}
  \caption{
    Reduced dynamics of a system TLS coupled to an environment TLS for an initial mixed joint state.
    Shown are the system populations and purity \(\Tr\left(\rho_S^2\right)\) obtained from the full unitary Liouville-von Neumann evolution (solid lines).
  }
  \label{fig:tls_tls_mixed}
\end{figure}

For mixed initial states, we evolve the density matrix in Liouville space~\cite{breuerPetruccioneTheoryOpenQuantum2010,cressmanSarpeshkarEmulationDensityMatrix2025}.

In the absence of dissipation, the density matrix evolves according to the Liouville-von Neumann equation,
\begin{equation}
  \hbar \frac{d\rho}{dt} = -i\left[H, \rho\right],
\end{equation}
which becomes, after vectorization,
\begin{equation}
  \hbar \odv{}{t}\flsket{\rho} = -i K_H \flsket{\rho},
\end{equation}
with
\begin{equation}
  K_H = I \otimes H - H^T \otimes I.
\end{equation}
With column stacking, the entries of \(\rho\) are grouped column by column.
Left multiplication by \(H\) acts within each column of \(\rho\), while right multiplication by \(H\) acts across each row.
Componentwise,
\begin{equation}
  \hbar \dot{\rho}_{ij}
  =
  \underbrace{-i\sum_k H_{ik}\rho_{kj}}_{\text{column \(j\) of \(\rho\)}}
  +
  \underbrace{i\sum_k \rho_{ik}H_{kj}}_{\text{row \(i\) of \(\rho\)}} .
\end{equation}
Thus, \(H\rho\) supplies the vertical, column-wise routing pattern, while \(\rho H\) supplies the horizontal, row-wise routing pattern with the opposite sign.
The Kronecker form is just the column-stacked representation of this rule:
\begin{equation}
  K_H
  =
  \underbrace{I\otimes H}_{\text{column-wise action}}
  -
  \underbrace{H^T\otimes I}_{\text{row-wise action}} .
\end{equation}
This is the same row-and-column current-routing rule used in the density matrix circuit construction of Ref.~\cite{cressmanSarpeshkarEmulationDensityMatrix2025}, now written in compact Liouville space form.
For a diagonal Hamiltonian, this reduces to the familiar coherent circuit admittance \(H_{ii}-H_{jj}\) on the \(\rho_{ij}\) coordinate, with zero coherent self-admittance on the density matrix diagonal.
Appendix~\ref{sec:app:liouville} summarizes the vectorization convention and the corresponding Schur complement reduction.

For any time-independent linear Liouville space generator \(K\) written in the form \(\hbar \odv{}{t}\flsket{\rho}=-iK\flsket{\rho}\), partitioning the density matrix coordinates into retained and eliminated sectors gives
\begin{equation}
  K =
  \begin{bmatrix}
    K_{PP} & K_{PQ} \\
    K_{QP} & K_{QQ}
  \end{bmatrix}.
\end{equation}
The retained density matrix coordinates obey
\begin{equation}
  \left(\hbar s I + iK_{PP} + \Sigma(s)\right) \flsket{\rho_P(s)}
  =
  \hbar \flsket{\rho_P(0)} + F(s),
\end{equation}
where
\begin{equation}
  \begin{aligned}
    \Sigma(s) & = K_{PQ} \left(\hbar s I + iK_{QQ}\right)^{-1} K_{QP},                     \\
    F(s)      & = -i\hbar K_{PQ} \left(\hbar s I + iK_{QQ}\right)^{-1} \flsket{\rho_Q(0)}.
  \end{aligned}
\end{equation}
For the chosen coordinate projection, this is the Laplace domain Nakajima-Zwanzig memory equation~\cite{nakajimaQuantumTheoryTransport1958,zwanzigEnsembleMethodTheory1960,breuerPetruccioneTheoryOpenQuantum2010}.
With \(\mathcal L=-iK/\hbar\), its memory kernel and initial-condition source satisfy \(\mathcal M(s)=-\Sigma(s)/\hbar\) and \(j(s)=F(s)/\hbar\).
The circuit construction realizes these terms as a dynamical admittance and an effective current source, with different choices of projectors producing different kernels.
For the coherent mixed state example, \(K=K_H\).
In Liouville space, the eliminated sector contains the other environment blocks as well as the cross-block coherences, so the poles of \(\Sigma(s)\) encode both population and coherence dynamics.

We compare the reduced Liouville space reconstruction with the exact full Liouville-von Neumann evolution for an initially mixed composite state.
Figure~\ref{fig:tls_tls_mixed} shows the resulting system populations and purity \(\Tr\left(\rho_S^2\right)\) are in agreement.

\subsection{\label{sec:composite:lindblad} Lindblad Dynamics in Liouville Space}

\begin{figure}
  \centering
  \includegraphics[max width=\columnwidth]{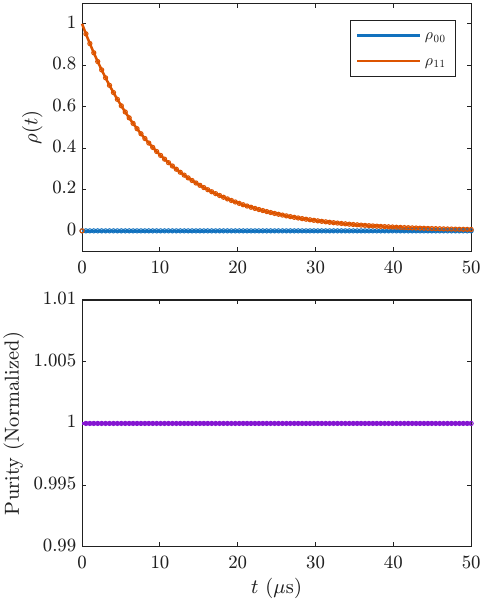}
  \caption{
    \(H_{\mathrm{eff}} = H - \frac{i\hbar}{2}L^\dagger L\) for the two-level decay channel \(L=\sqrt{\gamma}\ketbra{0}{1}\), initialized in the upper state.
    \emph{Top:} unnormalized diagonal probabilities under the effective Hamiltonian evolution.
    When the jump term is omitted, probability is not transferred into the lower state.
    \emph{Bottom:} normalized conditional purity \(\Tr(\rho^2)/[\Tr(\rho)]^2\), which remains unity because the conditioned state stays pure under this diagonal \(H_{\mathrm{eff}}\).
    Solid lines show the exact effective Hamiltonian evolution, and markers show the circuit simulation results.
  }
  \label{fig:tls_effective_hamiltonian}
\end{figure}

\begin{figure}
  \centering
  \includegraphics[max width=\columnwidth]{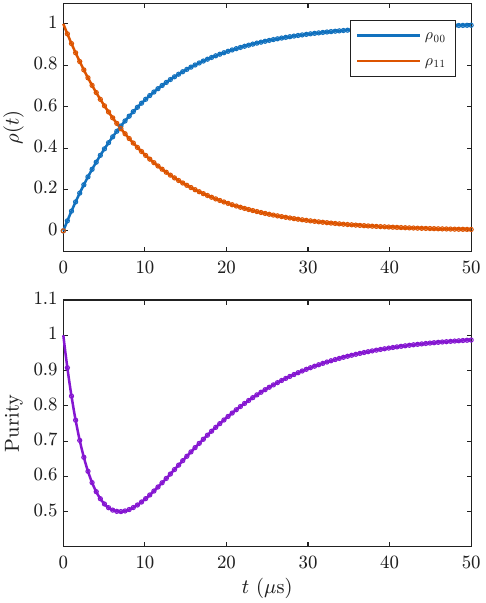}
  \caption{
    Full Lindblad evolution for the same two-level decay channel.
    \emph{Top:} trace-preserving population transfer from the upper state into the lower state through the effective Hamiltonian and jump term.
    \emph{Bottom:} the purity \(\Tr(\rho^2)\) of the normalized density matrix, initially unity, is lowest near equal populations and approaches unity again as the system relaxes into the lower state.
    Solid lines show the exact Lindblad evolution and markers show the circuit simulation results.
  }
  \label{fig:tls_lindblad}
\end{figure}

For a single decay channel, the density matrix evolves as~\cite{manzanoShortIntroductionLindblad2020,breuerPetruccioneTheoryOpenQuantum2010}
\begin{equation}
  \hbar \frac{d\rho}{dt}
  =
  -i[H,\rho]
  +
  \hbar L \rho L^\dagger
  -
  \frac{\hbar}{2}\left\{L^\dagger L,\rho\right\},
\end{equation}
where \(L\) is the jump operator defining the decay channel.

Using the column-stacking convention summarized in Appendix~\ref{sec:app:liouville}, the dynamics take the linear form
\begin{equation}
  \hbar \odv{}{t}\flsket{\rho} = -i K_{\mathrm{L}} \flsket{\rho},
\end{equation}
where
\begin{equation}
  \begin{aligned}
    K_{\mathrm{L}}    & = K_H + K_\mathrm{nojump} + K_\mathrm{jump},                                                    \\
    K_H               & = I \otimes H - H^T \otimes I,                                                                  \\
    K_\mathrm{nojump} & = -\frac{i\hbar}{2} \left(I \otimes L^\dagger L + \left(L^\dagger L\right)^T \otimes I \right), \\
    K_\mathrm{jump}   & = i\hbar \left(L^* \otimes L\right).                                                            \\
  \end{aligned}
\end{equation}
Here \(K_H\) gives the coherent Hamiltonian contribution.
The dissipative part is split into the non-Hermitian no-jump drift \(K_\mathrm{nojump}\) and the recycling term \(K_\mathrm{jump}\).
Together these terms produce trace-preserving Lindblad evolution.

The Norton reduction applies directly to dissipative Lindblad dynamics by taking \(K=K_{\mathrm L}\) in the Liouville space reduction above.
The Schur complement algebra is unchanged, but the Hamiltonian generator \(K_H\) is replaced by the generally non-Hermitian Lindblad generator \(K_{\mathrm L}\).
The following example shows the circuit contributions of no-jump loss and jump recycling.

We reorganize the dynamics in terms of the effective non-Hermitian Hamiltonian,
\begin{equation}
  H_{\mathrm{eff}} = H - \frac{i\hbar}{2}L^\dagger L,
\end{equation}
so that
\begin{equation}
  \begin{aligned}
    \hbar \odv{}{t}\flsket{\rho} & = -i \left(K_\mathrm{eff} + K_\mathrm{jump}  \right)\flsket{\rho}, \\
    K_\mathrm{eff}               & = I \otimes H_{\mathrm{eff}} - H_{\mathrm{eff}}^* \otimes I.
  \end{aligned}
\end{equation}
The no-jump evolution generated by \(H_{\mathrm{eff}}\) is not trace preserving, so in that sector we monitor the normalized conditional purity \(\Tr(\rho^2)/[\Tr(\rho)]^2\).
For the trace-preserving full Lindblad evolution we use the ordinary purity \(\Tr(\rho^2)\).

Because \(K_{\mathrm L}\) is generally non-Hermitian, the reduced Lindblad self-energy can contain poles away from the imaginary axis.
Appendix~\ref{sec:app:lindblad_complex_poles} gives circuit realizations of these complex-pole terms using auxiliary Planck capacitor pairs and voltage-controlled current sources.
Each subcircuit implements the pole dynamics and supplies a Norton output current weighted by the corresponding residue.

For the minimal two-level decay example, we take a diagonal two-level Hamiltonian and a single decay channel,
\begin{equation}
  H =
  \begin{bmatrix}
    E_0 & 0   \\
    0   & E_1
  \end{bmatrix},
  \qquad L =
  \sqrt{\gamma}\ketbra{0}{1},
\end{equation}
where \(\ket{1}\) denotes the decaying upper level and \(\ket{0}\) the lower level.
This yields
\begin{equation}
  H_{\mathrm{eff}} =
  \begin{bmatrix}
    E_0 & 0                             \\
    0   & E_1 - \dfrac{i\hbar\gamma}{2}
  \end{bmatrix}.
\end{equation}
Figures~\ref{fig:tls_effective_hamiltonian} and \ref{fig:tls_lindblad} separate the no-jump dynamics from the full dissipative Lindblad dynamics.
The effective Hamiltonian evolution in Fig.~\ref{fig:tls_effective_hamiltonian} maintains a constant normalized conditional purity because the total probability decays without any transfer of probability into the lower state.
By contrast, the full Lindblad evolution in Fig.~\ref{fig:tls_lindblad} includes the jump term and therefore the purity decreases through a transient mixed state regime before returning to unity as the system relaxes to the lower state.
\section{\label{conclusion} Discussion and Conclusion}

We have established subsystem elimination in quantum dynamics as an exact network reduction.
For a partitioned finite-dimensional time-independent Hamiltonian, eliminating the \(Q\) sector yields a closed equation on the retained sector in which the eliminated degrees of freedom appear as a dynamical load \(\Sigma(s)\) and a source term \(F(s)\).
In the single-port case this is obtained directly by Gaussian elimination, and in the general case it is the Schur complement of the eliminated block.
Under the circuit mapping, the eliminated subsystem is replaced by an exact Norton equivalent.

For finite eliminated sectors, \(\Sigma(s)\) is a rational function whose poles are fixed by the spectrum of the eliminated block and whose residues define how strongly those modes couple back to the retained coordinates.
This expansion connects memory to circuit dynamics and allows for approximations of the response with fewer modes.

The two-level and chain examples illustrate the progression from isolated poles to continuous pole spectra.
The Norton reduction also applies to density matrix elements, as demonstrated by the composite system and Liouville space examples, including trace-preserving open system evolution governed by the Lindblad equation.
In the effective Hamiltonian example, dissipation changes the pole locations but leaves the reduction procedure unchanged.

In the disorder-free Grover example, the marked state couples to the entire unmarked sector through a single bright pole, so the reduced model exposes the search dynamics as an exact collective branch.
With diagonal disorder, the reduced self-energy shows immediately how that structure changes: spectral weight is transferred out of the bright pole and into dark poles that are silent in the clean limit.
Circuit simulations show that the renormalized bright-only model accurately reproduces the marked state probability for weak disorder and captures its broad trend with increasing disorder.

Future work could use moment matching, Pad\'e-type approximations, and related model reduction tools to replace dense pole structure by smaller effective networks tailored to specific observables~\cite{antoulasApproximationLargescaleDynamical2005}.
Electrical noise and dissipative circuit elements could model stochastic and thermal environments with appropriately chosen statistics and couplings.
Finally, the finite-dimensional Lindblad constructions developed here could be extended to broader classes of open system reductions, drawing on recent exact reduction results for Lindblad dynamics~\cite{grigolettoViolaExactModelReduction2025}.
 
\begin{acknowledgments}
  This work was supported by Dartmouth College.
  We thank Douglas R. Beahm for early discussions and for assistance with adapting previously developed code for the initial stages of this work.
\end{acknowledgments}

\appendix

\section{\label{sec:app:inhomogeneous_norton} Norton Reduction of the Inhomogeneous Schr\"odinger Equation}

Inhomogeneous Schr\"odinger equations appear in a number of contexts where projector methods are used, such as in optimal control and reactive scattering~\cite{ndongKochChebychevPropagatorInhomogeneous2009}, as well as perturbation methods~\cite{vandijkToyamaNumericalSolutionsSchrodinger2014}.
The Schur complement reduction used for the homogeneous Schr\"odinger equation introduces a source term based on the initial condition of the eliminated sector, but the reduction can just as easily be used with a prescribed external time-dependent source.

Consider
\begin{equation}
  i \hbar \odv{}{t} \ket{\psi(t)} = H\ket{\psi(t)} + \ket{\eta(t)}
\end{equation}
where \(H\) is time-independent, \(\ket{\psi(t)}\) is a state or partial state that may not be normalized, and \(\ket{\eta(t)}\) is an external source independent of \(\ket{\psi(t)}\).
It is worth noting that the source \(\ket{\eta(t)}\) is distinct from Hamiltonian noise, which enters through random operators acting on the state.
Time-dependent Hamiltonians are addressed in Appendix~\ref{sec:app:time_dependent}.

Taking the unilateral Laplace transform gives
\begin{equation}
  \left(\hbar s I + i H\right) \ket{\Psi(s)} = \hbar \ket{\psi(0)} - i \ket{\eta(s)}
\end{equation}
Partitioning into coordinates \(P\) and \(Q\) and reducing gives us
\begin{equation}
  \begin{aligned}
    \left[\hbar s I + i H_{PP} + H_{PQ} \left(\hbar s I + i H_{QQ}\right)^{-1} H_{QP} \right]\Psi_P(s) \\
    = \hbar \psi_P(0) - i \hbar H_{PQ} \left( \hbar s I + i H_{QQ} \right)^{-1} \psi_Q(0)              \\
    - i \eta_P(s) - H_{PQ} \left( \hbar s I + i H_{QQ} \right)^{-1} \eta_Q(s).
  \end{aligned}
  \label{eq:norton_driven}
\end{equation}
The right-hand side presents three source terms (other than the initial condition \(\psi_P(0)\)) corresponding to the initial condition of the eliminated sector \(Q\) (as before) as well as the separate contributions from the \(P\) and \(Q\) sectors of the original source term \(\eta\).
Without loss of generality we may combine all three terms into a single \(F(s)\).

Thus, the inhomogeneous input \(\eta\) does not affect the determination of the reduced Norton admittance \(Y_\Sigma(s)\), only the Norton current source \(I_F(s)\).
\section{\label{sec:app:time_dependent} Norton Reduction with Time-Dependent Hamiltonians}

Taking the unilateral Laplace transform of the Schr\"odinger equation with a  time-dependent Hamiltonian generally leads to complicated expressions involving convolutions or shifted arguments in the Laplace variable \(s\).
However, simple substitutions in the time-domain still produce expressions for the quantum Norton equivalent admittance kernel \(Y_\Sigma(t,t')\) (which accounts for memory) and source \(I_F(t)\).
The time-independent formulation is trivially recovered from this formulation.

Consider a finite-dimensional Hermitian Hamiltonian \(H(t)\) with fixed retained subspace \(P\) and eliminated subspace \(Q\), as before.
It is convenient to define the time propagator of the isolated \(Q\) sector
\begin{equation}
    \begin{aligned}
        U_Q(t,t') \ket{\psi_Q^{(0)}(t')} & = \ket{\psi_Q^{(0)}(t)},                                                        \\
        U_Q(t,t')                        & = \mathcal{T} \exp{\left[-\frac{i}{\hbar} \int_{t'}^{t} H_{QQ}(u) \, du\right]}
    \end{aligned}
\end{equation}
where \(\ket{\psi_Q^{(0)}(t)}\) is the isolated, time-evolved \(Q\)-sector state vector and \(\mathcal{T}\) is the usual time-ordering operator~\cite{breuerPetruccioneTheoryOpenQuantum2010}.
Note that \(U_Q\) satisfies \(U_Q(t',t') = I\) and \(i\hbar \partial_t U_Q(t,t') = H_{QQ}(t) U_Q(t,t')\).

The solution of the eliminated sector \(Q\) (still coupled to \(P\)) may then be written as
\begin{equation}
    \begin{aligned}
        \psi_Q(t) & = U_Q(t,0) \psi_Q(0)                                                   \\
                  & - \frac{i}{\hbar} \int_{0}^{t} U_Q(t,t') H_{QP}(t') \psi_P(t') \, dt'.
    \end{aligned}
\end{equation}
Substitution then gives us an integro-differential equation for the retained sector \(P\):
\begin{equation}
    \hbar \odv{}{t} \psi_P(t) + i H_{PP}(t)\psi_P(t) + \int_{0}^{t} K_\Sigma(t,t') \psi_P(t') \, dt' = F(t)
\end{equation}
with
\begin{equation}
    \begin{aligned}
        F(t)           & = -i H_{PQ}(t) U_Q(t,0) \psi_Q(0)                \\
                       & = -i  H_{PQ}(t) \psi_Q^{(0)}(t),                 \\
        K_\Sigma(t,t') & = \frac{1}{\hbar} H_{PQ}(t) U_Q(t,t') H_{QP}(t')
    \end{aligned}
\end{equation}
where \(\psi_Q^{(0)}(t)\) is the evolution of the isolated \(Q\)-sector state.

The circuit mapping then is
\begin{equation}
    C_\hbar \odv{}{t} V_P(t) + Y_{PP}(t) V_P(t) + \int_{0}^{t} Y_\Sigma(t,t') V_P(t') \, dt' = I_F(t)
\end{equation}
where the current source term \(I_F(t)\) retains the same form as the time-independent Hamiltonian case: isolated \(Q\)-sector evolution with a (now) time-dependent coupling corresponding to \(H_{PQ}(t)\).
For the case of time-independent \(H_{QQ}\) (still with time-dependent \(H_{PQ}(t) = H_{QP}^\dagger(t)\)), we may simplify the memory kernel \(Y_\Sigma(t,t')\) and reuse the circuit structure presented in Fig.~\ref{fig:single_pole_realization} for each auxiliary mode of \(H_{QQ}\), with additional time-dependent input/output gains on the quantum transadmittance mixers.
Thus, for fixed \(H_{QQ}\), the existing Norton circuit construction extends to time-dependent Hamiltonians through modulation of the input and output coupling gains.
\section{\label{sec:app:finite_chain} Finite Chain Pole Expansion}

We consider a finite tight-binding chain of length \(L+1\) with open boundary conditions.
The first site is retained, while the remaining \(L\) sites form the eliminated subsystem \(H_{QQ}\).
The Hamiltonian \(H_{QQ}\) is diagonalized by a set of normal modes corresponding to standing wave solutions of the discrete Laplacian with open boundaries~\cite{ben-avrahamTamonOnedimensionalContinuoustimeQuantum2004,economouGreensFunctionsQuantum2006}:
\begin{equation}
  \begin{aligned}
    E_k       & = 2J \cos\left(\frac{\pi k}{L+1}\right),                    \\
    \phi_k(n) & = \sqrt{\frac{2}{L+1}}\sin\left(\frac{\pi k n}{L+1}\right),
  \end{aligned}
\end{equation}
for \(k = 1, \dots, L\), where \(n\) labels the site index within the eliminated \(Q\)-subspace.

The resolvent of the eliminated subsystem may be written in spectral form as
\begin{equation}
  \left(\hbar s I + iH_{QQ}\right)^{-1} = \sum_{k=1}^{L} \frac{\Pi_k}{\hbar s + iE_k},
\end{equation}
where \(\Pi_k = \proj{\phi_k}\) are the spectral projectors.

If the retained coordinate couples to the first site of the chain (i.e., \(n=1\) at the boundary between \(P\) and \(Q\)), the coupling matrices take the form
\begin{equation}
  H_{PQ} =
  \begin{bmatrix}
    J & 0 & \cdots & 0
  \end{bmatrix} , \quad H_{QP} = H_{PQ}^\dagger.
\end{equation}
The self-energy is then
\begin{equation}
  \begin{aligned}
    \Sigma_L(s) & = H_{PQ}\left(\hbar s I + iH_{QQ}\right)^{-1} H_{QP} \\
                & = \sum_{k=1}^{L} \frac{R_k}{\hbar s + iE_k},
  \end{aligned}
\end{equation} with residues \(R_k = H_{PQ} \Pi_k H_{QP}\), which reduce to scalars in this case since the retained subspace is one-dimensional.
Evaluating this expression gives
\begin{equation}
  \begin{aligned}
    R_k & = J^2 \vab{\phi_k(1)}^2                                  \\
        & = \frac{2J^2}{L+1} \sin^2\left(\frac{\pi k}{L+1}\right),
  \end{aligned}
\end{equation} which shows that each pole is weighted by the amplitude of the corresponding mode at the boundary site.
This expression is the discrete pole expansion of the boundary Green's function, whose continuum limit yields the semi-infinite chain response discussed in Sec.~\ref{sec:pole_structure:semiinfinite} and Appendix~\ref{sec:app:infinite_chain}.
\section{\label{sec:app:infinite_chain} Semi-Infinite Chain Time-Domain Solution}

For a continuous-time quantum walk on an infinite line with nearest-neighbor hopping Hamiltonian \(H = J\sum_n (\ketbra{n}{n+1} + \ketbra{n+1}{n})\), the propagator is given by~\cite{ben-avrahamTamonOnedimensionalContinuoustimeQuantum2004}
\begin{equation}
  K_m(t) = (-i)^m J_m\left(\frac{2Jt}{\hbar}\right),
\end{equation}
where \(J_m\) denotes the Bessel function of the first kind.

To obtain the semi-infinite chain, we impose a hard-wall boundary at \(n=0\) using the method of images.
We consider an initial condition localized at the boundary, \(\psi_n(0) = \delta_{n,1}\).
The resulting wavefunction is
\begin{equation}
  \psi_n(t) = K_{n-1}(t) - K_{n+1}(t).
\end{equation}

At the boundary site (\(n=1\)), this reduces to
\begin{equation}
  \psi_1(t) = J_0\left(\frac{2Jt}{\hbar}\right) + J_2\left(\frac{2Jt}{\hbar}\right).
\end{equation}

This time-domain expression corresponds to the continuum limit of the pole expansion described in Sec.~\ref{sec:pole_structure:semiinfinite}, where the discrete set of modes becomes a continuous energy band and the self-energy acquires a branch-cut structure associated with this continuum.
\section{\label{sec:app:grover_bright} Bright-pole reduction and approximations to the dark cloud}

This appendix develops the perturbative bright-pole reduction used in Sec.~\ref{sec:disordered_grover}.

\subsection{\label{sec:app:grover_bright:clean} Bright and dark modes of the clean \(Q\) sector}

We adopt the clean Grover setup of Sec.~\ref{sec:disordered_grover}, with
\begin{equation}
  H_{QQ}^{(0)}=\gamma(J_{N-1}-I_{N-1}),
  \qquad
  H_{PQ}=\gamma\sqrt{N-1}\bra{\psi_r},
\end{equation}
where \(\ket{\psi_r}\) is the symmetric unmarked state on the \(Q\) sector and \(J_{N-1}\) denotes the \((N-1)\times(N-1)\) all-ones matrix.
Because \(H_{PQ}\) is proportional to \(\bra{\psi_r}\), the bright state is the unique \(Q\)-sector direction visible from the retained coordinate, and the dark sector is its orthogonal complement.
In the clean problem, this coupling-defined bright/dark split is also the eigenmode decomposition of \(H_{QQ}^{(0)}\).

Since \(J_{N-1}\ket{\psi_r}=(N-1)\ket{\psi_r}\),
\begin{equation}
  H_{QQ}^{(0)}\ket{\psi_r}=(N-2)\gamma\ket{\psi_r},
\end{equation}
so \(\ket{\psi_r}\) is the unique bright eigenmode with eigenvalue \(E^{(0)}_b=(N-2)\gamma\).
For any dark state \(\ket{\psi_d}\) with \(\braket{\psi_r}{\psi_d}=0\), the components of \(\ket{\psi_d}\) sum to zero, hence \(J_{N-1}\ket{\psi_d}=0\) and
\begin{equation}
  H_{QQ}^{(0)}\ket{\psi_d}=-\gamma\,\ket{\psi_d}.
\end{equation}
The orthogonal complement of \(\ket{\psi_r}\) is thus an \((N-2)\)-dimensional degenerate dark subspace with common eigenvalue \(E^{(0)}_d=-\gamma\).

The residues follow from the spectral projectors.
For \(\Pi^{(0)}_b=\proj{\psi_r}\),
\begin{equation}
  R^{(b)}_0 = H_{PQ}\Pi^{(0)}_bH_{QP} = (N-1)\gamma^2,
\end{equation}
while for any clean dark projector \(\Pi^{(0)}_{d,a}=\proj{\psi_{d_a}}\),
\begin{equation}
  R^{(0)}_{d,a} = (N-1)\gamma^2\vab{\braket{\psi_r}{\psi_{d_a}}}^2 = 0,
\end{equation}
since \(\ket{\psi_{d_a}}\) is orthogonal to \(\ket{\psi_r}\).
The dark manifold is therefore spectrally present in \(H_{QQ}^{(0)}\) but silent at the retained port, which is the derivation behind the clean single-pole \(\Sigma_{\mathrm{clean}}(s)\) and \(F_{\mathrm{clean}}(s)\) of Sec.~\ref{sec:disordered_grover}.

\subsection{\label{sec:app:grover_bright:perturbative} Perturbative bright-pole reduction}

We now restore the centered diagonal disorder of Sec.~\ref{sec:disordered_grover}, splitting the \(Q\) block as
\begin{equation}
  H_{QQ} = H_{QQ}^{(0)} + V_Q,
  \qquad
  V_Q = \sum_{j=1}^{N-1}\delta_{Q_j}\proj{\psi_{Q_j}},
\end{equation}
with \(\sum_{j=1}^{N-1}\delta_{Q_j}=0\) and \(\sigma,\tilde{\sigma}\) as defined in the main text.
Completeness of the \(Q\)-sector spectral projectors fixes the total residue independently of disorder,
\begin{equation}
  \begin{aligned}
    \sum_k R_k & = H_{PQ}\left(\sum_k \Pi_k\right)H_{QP}                 \\
               & = H_{PQ}H_{QP} = (N-1)\gamma^2 \equiv R_{\mathrm{tot}},
  \end{aligned}
\end{equation}
Thus \(V_Q\) changes the energies and redistributes a fixed total pole weight.
Whatever weight leaves the bright pole reappears in the dark cloud.

Centering also kills the first-order bright shift.
Since
\begin{equation}
  \braket[3]{\psi_r}{V_Q}{\psi_r} = \frac{1}{N-1}\sum_{j=1}^{N-1} \delta_{Q_j} = 0,
\end{equation}
the first-order bright shift vanishes:
\begin{equation}
  \Delta E^{(1)}_b = 0.
\end{equation}
The dark manifold is nevertheless lifted at first order by the projected perturbation \(\Pi_d^{(0)}V_Q\Pi_d^{(0)}\).
These first-order dark sector splittings do not modify the retained bright-pole formulas before \(\mathcal{O}(V_Q^3)\).

The leading bright-energy correction therefore arises at second order through virtual mixing with the dark subspace.
Using standard time-independent perturbation theory,\cite{sakuraiNapolitanoModernQuantumMechanics2021}
\begin{equation}
  \Delta E^{(2)}_b = \sum_{d\in\mathrm{dark}} \frac{\vab{\braket[3]{\psi_d}{V_Q}{\psi_r}}^2}{E^{(0)}_b-E^{(0)}_d}.
\end{equation}
Since the denominator is constant and equal to \((N-1)\gamma\), one obtains
\begin{equation}
  \begin{aligned}
    \Delta E^{(2)}_b
     & = \frac{1}{(N-1)\gamma}
    \sum_{d\in\mathrm{dark}}
    \vab{\braket[3]{\psi_d}{V_Q}{\psi_r}}^2 \\
     & = \frac{1}{(N-1)\gamma}
    \braket[3]{\psi_r}{V_Q \Pi^{(0)}_d V_Q}{\psi_r},
  \end{aligned}
\end{equation}
where
\begin{equation}
  \Pi^{(0)}_d = I-\proj{\psi_r}
\end{equation}
is the clean dark projector.
Using
\begin{equation}
  \begin{aligned}
    \braket[3]{\psi_r}{V_Q \Pi^{(0)}_d V_Q}{\psi_r}
     & = \braket[3]{\psi_r}{V_Q^2}{\psi_r} - \left(\braket[3]{\psi_r}{V_Q}{\psi_r}\right)^2 \\
     & = \braket[3]{\psi_r}{V_Q^2}{\psi_r}                                                  \\
     & = \frac{1}{N-1}\sum_{j=1}^{N-1} \delta_{Q_j}^2,
  \end{aligned}
\end{equation}
the bright-pole energy becomes
\begin{equation}
  \begin{aligned}
    E^{(b)} & \approx \left(N-2\right)\gamma + \frac{1}{\left(N-1\right)^2\gamma} \sum_{j=1}^{N-1} \delta_{Q_j}^2 \\ &= \left(N-2\right)\gamma + \frac{\sigma^2}{\left(N-1\right)\gamma} \\ &= \left(N-2\right)\gamma + \gamma \tilde{\sigma}^2.
  \end{aligned}
\end{equation}

The bright eigenvalue has no first-order shift, but the bright eigenvector acquires a first-order dark admixture.
The corresponding normalized corrected bright state is
\begin{equation}
  \ket*{\psi_r^{(b)}}
  \approx
  \left(1-\frac{1}{2}\eta^2\right)\ket{\psi_r}
  +
  \sum_{d\in\mathrm{dark}}
  \frac{\braket[3]{\psi_d}{V_Q}{\psi_r}}{(N-1)\gamma}\ket{\psi_d},
\end{equation}
with
\begin{equation}
  \begin{aligned}
    \eta^2 & = \sum_{d\in\mathrm{dark}} \vab{\frac{\braket[3]{\psi_d}{V_Q}{\psi_r}}{\left(N-1\right)\gamma}}^2 \\
           & = \frac{1}{\left(N-1\right)^3\gamma^2} \sum_{j=1}^{N-1} \delta_{Q_j}^2                            \\
           & = \frac{\sigma^2}{\left(N-1\right)^2\gamma^2}                                                     \\
           & = \frac{\tilde{\sigma}^2}{N-1}.
  \end{aligned}
\end{equation}
Since the marked coordinate couples to the \(Q\) sector through \(H_{PQ}\) and \(H_{QP}\), the bright residue is
\begin{equation}
  \begin{aligned}
    R^{(b)} & = H_{PQ}\ketbra*{\psi_r^{(b)}}{\psi_r^{(b)}}H_{QP}       \\
            & \approx (N-1)\gamma^2 \left(1-\frac{1}{2}\eta^2\right)^2 \\
            & \approx (N-1)\gamma^2\left(1-\eta^2\right),
  \end{aligned}
\end{equation}
where the final line retains terms through \(\mathcal{O}(V_Q^2)\).

Substituting the perturbative bright-pole quantities gives
\begin{equation}
  \begin{aligned}
    \Sigma(s) & \approx \frac{\left(N-1\right)\gamma^2\left(1-\eta^2\right)}{\hbar s+i\left[\left(N-2\right)\gamma+\gamma\tilde{\sigma}^2\right]},                         \\
    F(s)      & \approx -i\hbar \frac{\left(N-1\right)\gamma}{\sqrt{N}} \frac{\left(1-\eta^2\right)}{\hbar s+i\left[\left(N-2\right)\gamma+\gamma\tilde{\sigma}^2\right]}.
  \end{aligned}
\end{equation}
These expressions define the bright-only reduced model used in Sec.~\ref{sec:disordered_grover}.

\subsection{\label{sec:app:grover_bright:approximations} Compact approximations to the dark cloud}

The bright-only model neglects the dark sector contribution
\begin{equation}
  \Sigma_d(s)=\sum_{m=1}^{N-2} \frac{R_m^{(d)}}{\hbar s + i E_m^{(d)}}.
\end{equation}
When this contribution significantly affects the marked state probability, one could approximate it with a smaller set of effective poles~\cite{antoulasApproximationLargescaleDynamical2005,grigolettoViolaExactModelReduction2025}.
A single effective pole can match the total dark sector residue and its weighted mean energy, while a second pole allows a nonzero energy spread to be represented.

More accurate approximations over a chosen frequency range may be constructed with standard rational-approximation tools such as Pad\'e approximation, vector fitting, or L\"oewner-based realization methods~\cite{gustavsenSemlyenRationalApproximationFrequency1999,gustavsenComputerCodeRational2002,mayoAntoulasFrameworkSolutionGeneralized2007}.
This approximation method is analogous to pseudomode constructions for open quantum systems, which represent a structured environment with a smaller set of effective modes~\cite{garrawayDecayAtomCoupled1997}.
\section{\label{sec:app:liouville} Liouville Space Reduction for Unitary and Dissipative Dynamics}

We work in the vectorized (Fock-Liouville) representation, in which both unitary and Lindblad dynamics become linear ordinary differential equations for the vectorized density matrix~\cite{breuerPetruccioneTheoryOpenQuantum2010,cressmanSarpeshkarEmulationDensityMatrix2025}.
We write
\begin{equation}
  \flsket{\rho} = \vecop\left(\rho\right),
\end{equation}
using the column-stacking convention
\begin{equation}
  \vecop\left(A \rho B\right) = \left(B^T \otimes A\right)\vecop\left(\rho\right).
\end{equation}
For a Hilbert space of dimension \(N\), this lifts the dynamics to dimension \(N^2\).

Direct application of this identity to the closed-system commutator \([H,\rho]\) produces the Hamiltonian Liouville space operator \(K_H=I\otimes H-H^T\otimes I\) of Sec.~\ref{sec:composite:mixed}.
Applied to the Lindblad dissipator, the same identity produces the operators \(K_{\mathrm{nojump}}\), \(K_{\mathrm{jump}}\), and the effective-Hamiltonian form \(K_{\mathrm{eff}}\) of Sec.~\ref{sec:composite:lindblad}.
The remainder of this appendix uses the generic energy-scaled Liouville space operator \(K\), standing for \(K_H\), \(K_{\mathrm{L}}\), \(K_{\mathrm{eff}}\), or \(K_{\mathrm{eff}}+K_{\mathrm{jump}}\), and derives the Schur complement reduction invoked in the main text.

For the composite-system reductions, suppose the Hilbert-space basis is written as \(\ket{e,s}\), where \(e\) labels the environment configuration and \(s\) labels the system coordinate.
The corresponding Liouville space basis is
\begin{equation}
  \vecop\left(\ketbra{e,s}{e',s'}\right).
\end{equation}
For the retained environment value \(e_0\), the natural retained coordinates are
\begin{equation}
  \rho^{(e_0)}_{ss'} = \rho_{(e_0,s),(e_0,s')}.
\end{equation}
For a two-level retained system block, this becomes
\begin{equation}
  \flsket{\rho_S^{(e_0)}} = \vecop\left(\rho_S^{(e_0)}\right) =
  \begin{bmatrix}
    \rho^{(e_0)}_{00} \\
    \rho^{(e_0)}_{10} \\
    \rho^{(e_0)}_{01} \\
    \rho^{(e_0)}_{11}
  \end{bmatrix}.
\end{equation}
Equivalently, the retained Liouville subspace is
\begin{equation}
  P = \operatorname{span} \left\{ \vecop\left(\ketbra{e_0,s}{e_0,s'}\right) \right\}_{s,s'},
\end{equation}
while \(Q\) contains all remaining coordinates, including the other environment blocks as well as coherences between blocks.

Taking the Laplace transform gives
\begin{equation}
  \left(\hbar s I + iK\right)\flsket{\rho(s)} = \hbar \flsket{\rho(0)}.
\end{equation}

Partitioning the Liouville space operator into retained and eliminated sectors,
\begin{equation}
  K =
  \begin{bmatrix}
    K_{PP} & K_{PQ} \\
    K_{QP} & K_{QQ}
  \end{bmatrix},
\end{equation}
the same Schur complement reduction yields
\begin{equation}
  \left(\hbar s I + iK_{PP} + \Sigma(s) \right)\flsket{\rho_P(s)}
  =
  \hbar \flsket{\rho_P(0)} + F(s),
\end{equation}
with
\begin{equation}
  \begin{aligned}
    \Sigma(s)
     & = K_{PQ}\left(\hbar s I + iK_{QQ}\right)^{-1} K_{QP},                      \\
    F(s)
     & = -i \hbar K_{PQ}\left(\hbar s I + iK_{QQ}\right)^{-1} \flsket{\rho_Q(0)}.
  \end{aligned}
\end{equation}

The structure is identical to the state vector case, with the Hamiltonian replaced by the appropriate Liouville space generator.
When an explicit environment basis is used, the retained coordinates are density matrix elements conditioned on the fixed environment value \(e_0\), rather than wavefunction amplitudes.
The reduced self-energy therefore describes excursions out of the retained \(e=e_0\) block into the complementary Liouville space sector and back.
Its poles are determined by the resolvent
\(\left(\hbar s I+iK_{QQ}\right)^{-1}\).
For diagonalizable \(K_{QQ}\), they occur at
\(s_\alpha=-i\kappa_\alpha/\hbar\), where \(\kappa_\alpha\) are the
eigenvalues of \(K_{QQ}\).
When \(K_{QQ}\) is non-Hermitian, as in the Lindblad case, these poles represent decaying and oscillatory Liouville space modes, and the corresponding right and left eigenvectors need not be orthogonal.
From a circuit perspective, the Liouville space lifting introduces additional dynamical branches associated with coherences.
\section{\label{sec:app:lindblad_complex_poles} Circuit Realization of Lindblad Memory Kernels with Complex Poles}

\begin{figure*}
  \centering
  \includegraphics[max width=\columnwidth]{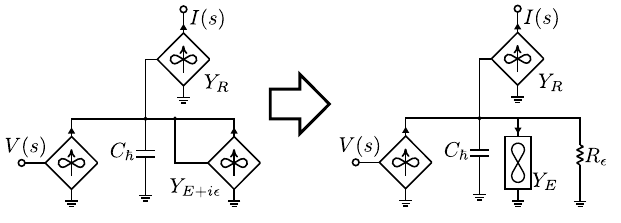}
  \caption{
    Schematic realization of a single complex-pole memory branch in the reduced Lindblad kernel.
    The retained input voltage \(V(s)\) drives an auxiliary branch through an input quantum transadmittance.
    The auxiliary branch state is stored as a voltage on the Planck capacitor \(C_{\hbar}\).
    A self-coupled quantum transadmittance \(Y_{E+i\epsilon}\) implements the local complex pole dynamics of the auxiliary branch, while a second transadmittance \(Y_R\) maps the auxiliary branch voltage to the returned Norton current \(I(s)\).
    Together these elements realize a branch of the form \(I(s)=\dfrac{R}{C_{\hbar}s-\epsilon+iE}V(s)\).
    For \(\epsilon<0\), the real pole shift is dissipative and can equivalently be implemented by a positive resistor \(R_\epsilon=-1/\epsilon\) in parallel with \(C_{\hbar}\), as shown on the right.
  }
  \label{fig:lindblad_complex_pole_schematic}
\end{figure*}

A Norton reduction may be carried out for the Lindblad generator in exactly the same Schur complement sense as for the Schr\"odinger and Liouville-von Neumann cases.
The difference is that eliminating Liouville space coordinates can produce reduced memory kernels with poles away from the imaginary axis, so the purely imaginary-pole circuit of Fig.~\ref{fig:single_pole_realization} must be extended to a complex-pole branch.

We use the quantum-domain convention
\begin{equation}
  \Sigma(s)=\frac{R}{\hbar s+i(E+i\epsilon)}=\frac{R}{\hbar s-\epsilon+iE}.
\end{equation}
Here \(R=R^R+iR^I\) is the residue, \(E\) is the oscillation energy, and \(\epsilon\) sets the real-part shift of the pole.

In the circuit realization, we identify \(\hbar \equiv C_{\hbar}\), so the corresponding branch is written directly in circuit variables.
The implementation takes the real and imaginary components of the amplitude of the retained sector \(V=V^R+iV^I\) and outputs a complex current \(I=I^R+iI^I\).

The subcircuit implements
\begin{equation}
  I(s)=\frac{R}{C_{\hbar}s-\epsilon+iE}V(s).
\end{equation}
This is done by introducing an internal complex state \(x=x^R+ix^I\) satisfying
\begin{equation}
  (C_{\hbar}s-\epsilon+iE)x(s)=V(s),
\end{equation}
followed by the output relation
\begin{equation}
  I(s)=R\,x(s).
\end{equation}
Thus, the pole location is set by the internal branch equation, while the residue is applied in the output current mapping.

Writing the pole equation in the time domain and separating real and imaginary parts gives
\begin{equation}
  C_{\hbar} \odv{}{t}
  \begin{bmatrix}
    x^R \\
    x^I
  \end{bmatrix}
  =
  \begin{bmatrix}
    \epsilon & E        \\
    -E       & \epsilon
  \end{bmatrix}
  \begin{bmatrix}
    x^R \\
    x^I
  \end{bmatrix}
  +
  \begin{bmatrix}
    1 & 0 \\
    0 & 1
  \end{bmatrix}
  \begin{bmatrix}
    V^R \\
    V^I
  \end{bmatrix}.
\end{equation}

The circuit implementation follows directly from these equations.
The retained input voltage \(V(s)\) drives an auxiliary branch through an input quantum transadmittance.
The auxiliary branch state is stored as a voltage on the Planck capacitor \(C_{\hbar}\).
A self-coupled quantum transadmittance implements the local complex-pole dynamics of the auxiliary branch, while a second transadmittance \(Y_R\) maps the auxiliary branch voltage to the returned Norton current \(I(s)\).
Together these elements realize a branch of the form \(I(s)=\dfrac{R}{C_{\hbar}s-\epsilon+iE}V(s)\).
Figure~\ref{fig:lindblad_complex_pole_schematic} shows the corresponding schematic realization.
\section{\label{sec:app:scaling} Energy, Power, and Part Scaling of Analog, Digital, Collective Analog, and Quantum Computers}

\begin{table*}[ht]
  \setlength{\extrarowheight}{4pt}
  \begin{tabularx}{\textwidth}{|X|X|}
    \hline
    \centering\textbf{Analog}
     & \centering\textbf{Digital} \tabularnewline
    \hline
    Function on a continuous set of numbers, e.g., \([0,1]\) or graded protein production from low to a maximum level.
     & Function on a discrete set, e.g., \(\{0,1\}\) or protein produced at a maximum level or not present at all.                                                                                            \\
    \hline
    The foundational functions arise from the physics and chemistry of the underlying devices such that the amount of function squeezed out of a single transistor, DNA, RNA, protein, neural, or quantum circuit is high.
     & The foundational functions arise from the mathematics of Boolean logic such that the amount of function squeezed out of a single transistor, genetic, RNA, protein, neural, or quantum circuit is low. \\
    \hline
    One wire or channel can represent many bits of information.
     & One wire or channel always represents one bit of information.                                                                                                                                          \\
    \hline
    Function is sensitive to the parameters of the circuits.
     & Function is less sensitive to the parameters of the circuits.                                                                                                                                          \\
    \hline
    Noise is due to thermal fluctuations in devices.
     & Noise is due to roundoff error and temporal aliasing.                                                                                                                                                  \\
    \hline
    Signal is not restored at each stage of a circuit.
     & Signal is restored at each stage of the circuit.                                                                                                                                                       \\
    \hline
    Robust at final and decisive outputs
     & Robust in every device and signal                                                                                                                                                                      \\
    \hline
    Not As Programmable But You Can Learn/Calibrate
     & Highly and Arbitrarily Programmable And Flexible And Scalable                                                                                                                                          \\
    \hline
  \end{tabularx}
  \caption{Comparison of analog and digital computation, adapted from Ref.~\cite{sarpeshkarAnalogSyntheticBiology2014}.}
  \label{tab:analog_digital_comparison}
\end{table*}

\begin{figure}
  \centering
  \includegraphics[max width=\columnwidth]{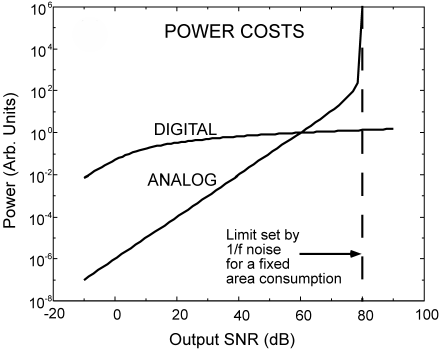}
  \caption{
    Analog and digital power costs versus output signal-to-noise ratio (SNR), adapted from Ref.~\cite{sarpeshkarUltraLowPower2010}.
    Power is shown in arbitrary units.
    The curves illustrate the crossover between analog and digital power costs and the limit imposed by \(1/f\) noise at fixed area consumption.
  }
  \label{fig:analog_digital_crossover}
\end{figure}

\begin{figure}
  \centering
  \includegraphics[max width=\columnwidth]{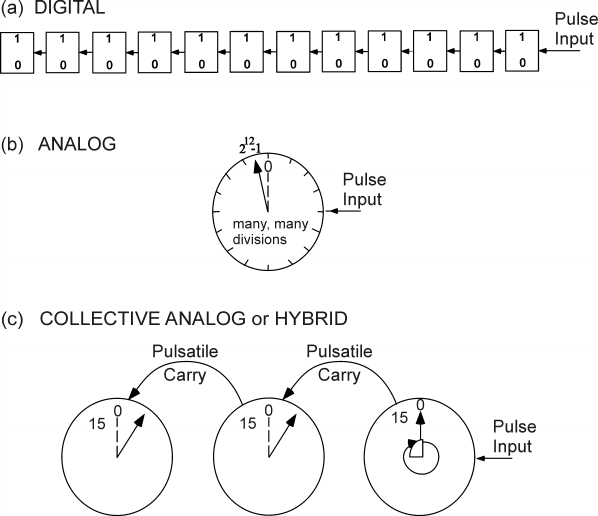}
  \caption{
    Pulse counting with (a) twelve one-bit digital units, (b) one twelve-bit analog unit, and (c) three four-bit collective analog units coupled by pulsatile carries.
    Each architecture detects a count exceeding \(2^{12}-1\).
    Adapted from Ref.~\cite{sarpeshkarUltraLowPower2010}.
  }
  \label{fig:collective_analog_counter}
\end{figure}

\begin{figure}
  \subfloat[]{
    \centering
    \includegraphics[max width=\columnwidth]{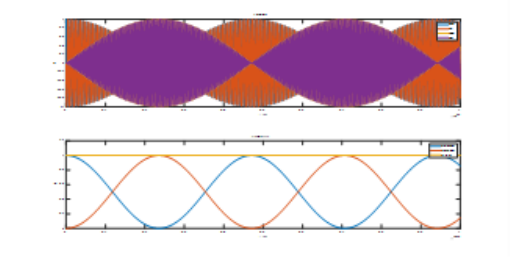}
  }
  \\
  \subfloat[]{
    \centering
    \includegraphics[max width=\columnwidth]{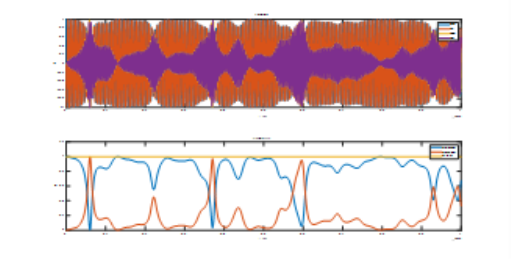}
  }
  \caption{
    Driven two-state dynamics without noise (top) and with illustrative spontaneous-emission noise (bottom).
    Each case shows state-variable trajectories and the corresponding probabilities.
  }
  \label{fig:driven_two_state_noise}
\end{figure}

\begin{figure}
  \centering
  \includegraphics[max width=\columnwidth]{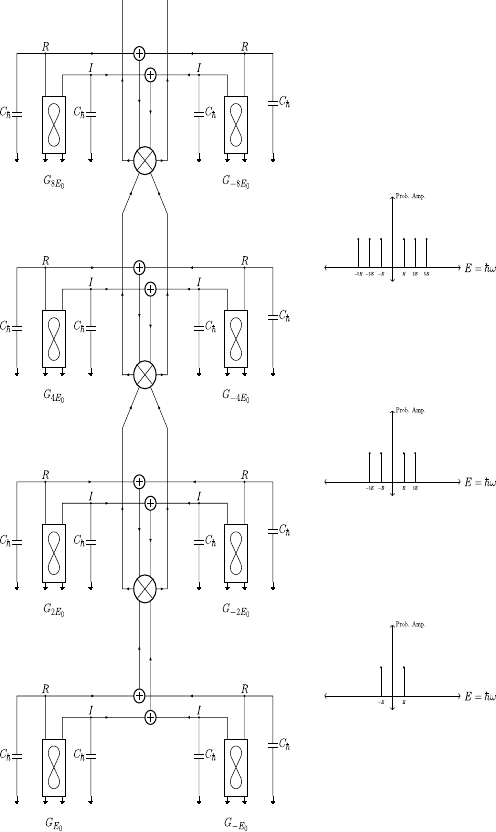}
  \caption{
    An architecture for the tensor product as a series of cascading complex multiplications.
    Adapted from Ref.~\cite{cressmanSarpeshkarFormulationEmulationQuantuminspired2022}.
  }
  \label{fig:tensor_product_architecture}
\end{figure}

\begin{figure}
  \centering
  \includegraphics[max width=\columnwidth]{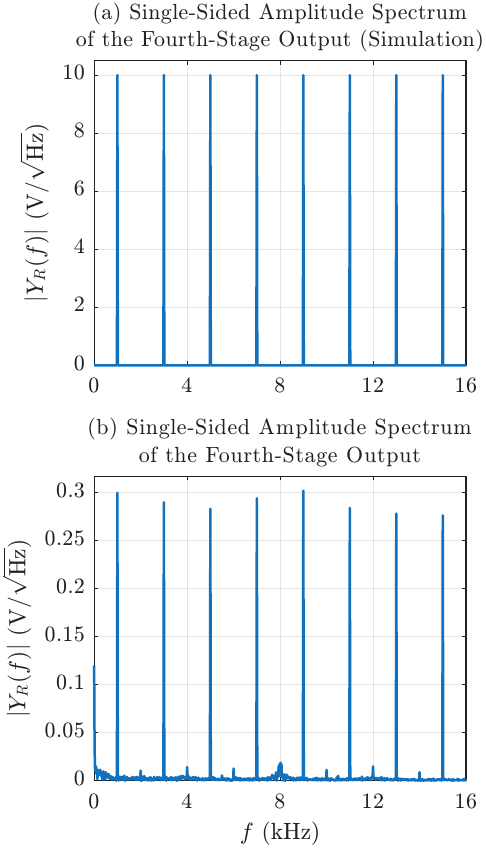}
  \caption{
    We compare the theoretical (a) and experimental (b) spectra of a simple tensor product input.
    Adapted from Ref.~\cite{cressmanSarpeshkarFormulationEmulationQuantuminspired2022}.
  }
  \label{fig:tensor_product_spectra}
\end{figure}

Table~\ref{tab:analog_digital_comparison} reviews robustness-efficiency tradeoffs between analog and digital computers~\cite{sarpeshkarAnalogSyntheticBiology2014}.
Analog and probabilistic digital computation is ubiquitous in physics (analog wave functions with probabilistic discrete outcomes after measurement in quantum physics and the continuous and statistical mathematics of classical physics), in chemistry (the quantum mechanical interactions of covalent bonds and the thermodynamic and classical interactions of other bonds), and in biology (reaction-diffusion equations in cells).

Mathematical analyses of the Poisson shot noise of thermally generated diffusive currents in electronic devices at room temperature, of noise in ionic and synaptic vesicle release and transport in neurobiology, of noise in the probabilistic and thermodynamic reaction-diffusion equations in cells, and of the Poisson noise of quantum atoms undergoing spontaneous emission in thermal equilibrium with their surroundings (as embodied in the Planck radiation formula, Lindblad equation, or the quantum fluctuation-dissipation theorem) lead to crossover curves similar to those in Fig.~\ref{fig:analog_digital_crossover}.

At relatively low precision, analog computation is highly efficient: It exploits basis functions for computation already built into nature with very efficient use of resources including energy/power, time, or space, e.g., exploiting Kirchhoff's current law to perform addition, rather than using many full-adder logic circuits with high overhead and device count, as in digital computation.
However, as computational precision gets higher, the costs of maintaining sufficient signal-to-noise ratio on a continuous analog variable and of implementing Kirchhoff's current law to high precision increase~\cite{sarpeshkarAnalogDigitalExtrapolating1998,sarpeshkarUltraLowPower2010}.
For example, in thermal noise-limited systems, it is \((2^8)^2\) times more expensive to implement 16-bit addition vs. 8-bit addition in analog systems~\cite{sarpeshkarUltraLowPower2010}.
In contrast, the divide-and-conquer collective strategy utilized by digital computation uses many 1-bit-precise local full-adder digital circuits and has them collectively interact via the mechanism of a digital carry.
It is scalable to arbitrary precision and efficient: it is only twice as costly to implement 16-bit-precise addition vs. 8-bit-precise addition, digitally.
Thus, as Fig.~\ref{fig:analog_digital_crossover} illustrates, there is crossover in power costs for analog versus digital and in collective analog and probabilistic computation with neurons in the brain~\cite{sarpeshkarAnalogDigitalExtrapolating1998,hahnloserSeungDigitalSelectionAnalogue2000,sarpeshkarUltraLowPower2010}.
The crossover can also be for other resources like area/part consumption.
Related crossover behavior has also been analyzed for computation in biological cells~\cite{sarpeshkarAnalogSyntheticBiology2014}.
The crossover location can be improved with feedback loops, calibration, error correction, and digitally-assisted machine learning.
However, a collective strategy is essential in ensuring that analog computation is scalable to arbitrary precision and complexity.
Quantum computers, which also exploit analog basis functions, also require collective analog or hybrid mixed-signal strategies for fault-tolerant computation.

A strategy that compares digital, analog, and collective analog hybrid computation is shown in Fig.~\ref{fig:collective_analog_counter}~\cite{wooSarpeshkarDigitallyProgrammableCytomorphic2018,sarpeshkarUltraLowPower2010}.
This example was chosen in part because it illustrates several features relevant to noise and errors in quantum computation as well.

Suppose we want to answer the following question: Is the number
of input pulses greater than \(2^{12}-1\)?
The pulses could be classical pulses, NMR \ang{90} pulses as in the previously demonstrated chip emulations, actual NMR pulses, spikes in neurobiology, or pulses in any other setting.
In all settings, a purely digital strategy could use 12 1-bit-precise toggling register units, with the pulse as an input clock to the least significant bit (LSB) unit on the right, and watch for an overflow at the most significant bit (MSB) unit on the left to signal `yes' as the answer.
The digital strategy is scalable to arbitrary precision but inefficient in its use of parts.
It has constant and high switching power dissipation.
More parts also imply higher overhead and higher leakage power dissipation.

A purely analog strategy could use one 12-bit-precise relatively large capacitor unit with each input pulse causing a small increment in a synaptic input charge on the capacitor; many small charge increments from successive input pulses can then effectively build charge on this capacitor.
At \(2^{12}-1\) input pulses, we are just below the threshold.
When the \(2^{12}\)th pulse arrives, we go above threshold, firing a pulsatile spike, and reset back to 0 (ground) to signal `yes'.
We are effectively doing `phase estimation' with the threshold being \(2\pi\) in phase~and \((2\pi-\epsilon)\) required to be a subthreshold phase, thus necessitating a low value of \(\epsilon\) (\(1/2^{12}\)) and high precision in the phase estimation.
The high precision forces us to use a relatively large capacitor (thermal reset noise scales like \(kT/C\) and mismatch in analog devices requires higher capacitance for matching as well).
It also forces us to use low-noise amplifiers to measure and spike with precision, while maintaining bandwidth, thus making the overall scheme power inefficient.

A collective analog or hybrid strategy is shown in Fig.~\ref{fig:collective_analog_counter}(c).
It uses three 4-bit-precise units (for example) such that any single capacitor/unit only needs 4 bits of precision for its charge increment and its precision (16 values to count before we overrange and reset); as in the digital strategy, the analog units interact collectively with the overranging spike in any one unit serving as a `carry' to its adjacent unit.
The latter unit can increment 1 unit for every 16 units of increment of the neighbor to its right.
Applying the idea recursively, three 4-bit-precise analog units using only Kirchhoff's current law and charge conservation to add can solve the problem with a more efficient use of parts, lower switching power dissipation, and lower overhead power dissipation.
Furthermore, intermediate analog values in any one unit need only be implicitly represented and never measured.
The measurement of the overranging final output digital pulse is all that is needed.
Such a collective analog adder that is scalable to arbitrary precision has been demonstrated on a working-hardware VLSI chip~\cite{woosungsikSarpeshkarSpikingneuronCollectiveAnalog2013}.
Although counting is a simple computational task, hybrid state machines capable of vector pattern recognition~\cite{sarpeshkarOHalloranScalableHybridComputation2002}, learning, arbitrary sequential computation with hybrid state machines and analog-to-digital conversion with spikes~\cite{sarpeshkarOHalloranScalableHybridComputation2002,yangSarpeshkarTimebasedEnergyefficientAnalogtodigital2005,yangSarpeshkarBioinspiredUltraenergyefficientAnalogtodigital2006} have also been demonstrated.
The analog-to-digital converter in Refs.~\cite{yangSarpeshkarTimebasedEnergyefficientAnalogtodigital2005,yangSarpeshkarBioinspiredUltraenergyefficientAnalogtodigital2006} was the first to show that \(2^n\) analog states (\(n\) bits) could be converted with only \(O(n)\) use of time resources rather than \(O(2^n)\).
It was the world's most energy efficient analog-to-digital converter at the time of publication.
As another example, ultra low power digitally-programmable cochlear implant processors for the deaf worked on subjects on the first try and led to breakthrough innovations~\cite{sarpeshkarEtAlUltralowpowerProgrammableAnalog2005,sitSarpeshkarCochlearimplantProcessorEncoding2008}.

Neurons in the human brain use spiking neuronal collective analog and probabilistic computation (synapses are highly noisy as are neurons) to implement pattern recognition and learning at high energy efficiency and with good fault tolerance: \qty{1}{\milli\second} neurons within the compact architecture of the brain, which are noisy and probabilistic, learn language with \(\approx 10^{2}\) fewer examples than a neural-network digital artificial intelligence (AI) machine consuming hundreds of \unit{\kilo\watt} of power.
The power consumption of data centers in the United States implementing probabilistic machine learning and such computations is already that of a nation state and growing rapidly.
Energy-efficient AI is thus critical.

We exploited Boltzmann thermodynamic physical laws governing noisy Poisson current flow in subthreshold transistors to show that collective analog computers could enable rapid slow-and-parallel stochastic computation with flat scaling in the number of reactions or molecules and milliwatt-scale power consumption~\cite{wooSarpeshkarDigitallyProgrammableCytomorphic2018,kimSarpeshkarFastPreciseEmulation2018}.
The latter chip implemented the Gillespie stochastic simulation algorithm (GSSA)~\cite{kimSarpeshkarFastPreciseEmulation2018}.
No digital supercomputer has been able to achieve a speedup of an exact GSSA implementation for over 50 years, regardless of power: Digital computers cannot parallelize such computations; furthermore, the high cost of random number generation and asynchronous Poisson noise scales prohibitively with the number of molecules or reactions.
Figure~\ref{fig:driven_two_state_noise} shows that we can use a similar strategy to simulate spontaneous emission in a driven time-dependent laser in our two-state systems as well (we show a very noisy system for illustration).
Finally, a key reason why the 2019 Google random-gate quantum chip substantially outperformed digital computers was not just because of quantum tensor product scaling that no digital computer can reproduce; it was also because it was capable of probabilistic measurement computations that no digital computer could do efficiently with pseudorandom number generation~\cite{aruteEtAlQuantumSupremacyUsing2019}.
The chip consumed nearly 26 kW of power and required refrigeration~\cite{aruteEtAlQuantumSupremacyUsing2019}.

The top panels of Fig.~\ref{fig:driven_two_state_noise} show non-noisy dynamics similar to those obtained from the previously demonstrated VLSI chip.
The bottom panels of Fig.~\ref{fig:driven_two_state_noise} incorporate such noise.
Section~\ref{sec:composite:lindblad} presents circuit simulations of two-level Lindblad decay (Fig.~\ref{fig:tls_lindblad}), and Appendix~\ref{sec:app:lindblad_complex_poles} gives circuit realizations of the associated memory kernels (Fig.~\ref{fig:lindblad_complex_pole_schematic}).

Quantum computers outperform digital computers for non-probabilistic reasons as well.
They use analog wave interference and superposition of an exponentially large number of possibilities via a tensor product in a large Hilbert space of dimension \(2^n\) with only \(O(n)\) qubit resources.
The previously demonstrated VLSI chip exploits this tensor product structure in a classical analog implementation.
Figures~\ref{fig:tensor_product_architecture} and~\ref{fig:tensor_product_spectra} show how successive quantum transadmittance mixers enable multiplicative operations in the time domain to yield convolved spectra in the Fourier domain that automatically implement favorable tensor product scaling.
The circuit architecture and experimental implementation are described in Ref.~\cite{cressmanSarpeshkarFormulationEmulationQuantuminspired2022}, with the fabricated chip and representative QFT results reproduced in Fig.~\ref{fig:chip_results}.
This tensor product architecture enabled the classical analog QFT implementation using \(O(n)\) circuit parts described in that work.

Classical analog emulations of actual quantum physics in nature are, in principle, less resource-efficient than actual quantum implementations for two key reasons: (1) Since we need paired real and imaginary variables (Fig.~\ref{fig:tls_mapping}), rather than one integrated complex variable as is available in nature, the component-count scaling of this implementation is always twice as costly with regard to the use of parts as in nature; (2) The four circuit primitives (Fig.~\ref{fig:circuit_symbols})~\cite{cressmanSarpeshkarFormulationEmulationQuantuminspired2022} are necessary and sufficient to emulate coherent quantum dynamics exactly: no LCR circuits can emulate the broadband \ang{-90} phase shift at all frequencies, required by the crucial \(i\) in the Schr\"odinger equation unless an infinite number of parts are used; no classical circuits can use single-element passive conservative real-voltage real-current devices with one voltage and one associated current to emulate complex voltage/current quantum dynamics.
That requires two voltages to be in quadrature and their corresponding two currents to be in anti-quadrature as we have implemented; finally, interactions between parts need to be modular and scalable and not subject to changing resonances as soon as they are wired together.

Thus, the paired real-imaginary variables, perfect broadband \ang{-90} shift, scalable broadband modularity, and two-voltage-two-current relationships necessitate active devices for classical analog circuits to emulate quantum mechanics.
Active devices, however, consume energy, unlike the passive devices of quantum mechanics.

Hence, we need exponentially more energy for \(2^n\) superpositions as we scale the energy values of the \(O(2N)\) circuit parts to ensure no aliasing in the spectrum by the Nyquist sampling theorem.
The component count must therefore be considered together with bandwidth, precision, and energy requirements.
Fortunately, as we have discussed, even 40 qubits from \qty{1}{\hertz} to \qty{10}{\giga\hertz} in modern silicon only require tens of \unit{\milli\watt} to implement.

Near-term quantum computers remain expensive, physically large systems that require substantial supporting infrastructure, including refrigeration.
Together with limitations in single- and two-qubit gate fidelity, these constraints suggest that widespread practical applications may remain years away and that quantum computing may be particularly well suited to centralized, cloud-based deployment.
Meanwhile, there are application spaces where it can be advantageous to explore classical analog emulations in three areas:
(1) chemistry and materials science that require both classical and quantum simulations;
(2) novel hybrid quantum and classical algorithms that leverage both biology and physics, e.g., the broadband wavelet transform-like quantum cochlea that effectively implements a time-dependent nonlinear Schr\"odinger wave-like equation with good spectral fidelity~\cite{sarpeshkarQuantumCochleaEfficient2019}.
It does so even with high levels of loss for broadband spectral analysis, expanding the range of quantum algorithms beyond just three (QFT-based, Grover, and quantum simulation);
(3) systems where portability and cost can benefit from large tensor product quantum-inspired search chips.

Therefore, we need a rigorous foundation for simulating, designing, and analyzing deterministic and stochastic quantum-and-classical systems that are simultaneously quantum and classical, portable, and simultaneously deterministic and stochastic.
Fortunately, the quantum Norton theorem establishes a foundation for how such design and analysis may be done with well-established analog circuit software ported to quantum systems.
Such software can help design actual quantum systems or quantum-inspired systems.
They can help practical, applied design techniques create working-hardware VLSI chips.
The framework presented here is intended to help fulfill these needs.

The parts that we have described are necessary and sufficient and work in a fashion that is useful for the design of well-known algorithms like Grover's search, as demonstrated by the reduced Grover dynamics in Sec.~\ref{sec:grover} and Fig.~\ref{fig:grover_reduced_plot}.
In both quantum and quantum-inspired analog computing, the Hamiltonian scaling of connectivity is identical for fully connected Hamiltonians or those with more local interactions, regardless of the number of qubits, with each complex coupling represented by up to four real coupling terms due to the paired real and imaginary variables.

In summary:
(1) Just as artificial neural networks do not require biological neurons, quantum-inspired computation does not necessarily require quantum hardware.
The mathematical properties underlying these approaches can be useful independently of their physical implementation.
Classical implementations of quantum-inspired computation may therefore offer advantages for energy-efficient and embedded computing, with potential applications in chemistry, materials science, and drug discovery.
(2) Analog and probabilistic computers include both quantum and classical computers, and the quantum Norton theorem is meant to help unify both these frameworks from a practical circuit viewpoint.
Practical computation requires careful consideration of energy, time, and space costs.
Nature inspires us to combine the two better as she already does in quantum and classical chemistry, in Boltzmann reaction-diffusion equations in cells, in cochleas, in birds' magnetic compasses, and in human brains!
 
\bibliography{refs}

\end{document}